\documentclass[amsmath,amssymb,aps,superscriptaddress,preprint]{revtex4}
\usepackage{chapterbib}
\usepackage{amsmath}
\usepackage{amssymb}
\usepackage{feynmp}
\usepackage{color}
\usepackage{graphics}
\usepackage{graphicx}
\usepackage[colorlinks=true]{hyperref}
\begin{document}
\unitlength=1mm
\title{Effective Action Approach for Quantum Phase Transitions in Bosonic Lattices}
\author{Barry Bradlyn}
\affiliation{Department of Physics, Massachusetts Institute of Technology, Cambridge, Massachusetts 02139, USA}
\email{bbradlyn@mit.edu}
\author{Francisco Ednilson A. dos Santos}
\affiliation{Institut f\"ur Theoretische Physik, Freie Universit\"at Berlin, Arnimallee 14, 14195 Berlin, Germany}
\email{santos@physik.fu-berlin.de}
\author{Axel Pelster}
\affiliation{Fachbereich Physik, Universit\"at Duisburg-Essen, Campus Duisburg, Lotharstrasse 1, 47048 Duisburg, Germany}
\email{axel.pelster@uni-duisburg-essen.de}
\date{\today}

\begin{abstract}
Based on standard field-theoretic considerations, we develop an effective action approach for investigating quantum phase 
transitions in lattice Bose systems at arbitrary temperature. 
We begin by adding to the Hamiltonian of interest a symmetry breaking source term. 
Using time-dependent perturbation theory, we then expand the grand-canonical free energy as a double power series in both the 
tunneling and the source term. 
From here, an order parameter field is introduced in the standard way, and the underlying effective action is derived via a 
Legendre transformation. 
Determining the Ginzburg-Landau expansion to first order in the tunneling term, expressions for the Mott insulator--superfluid 
phase boundary, condensate density, average particle number, and compressibility are derived and analyzed in detail. 
Additionally, excitation spectra in the ordered phase are found by considering both longitudinal and transverse variations of the 
order parameter. 
Finally, these results are applied to the concrete case of the Bose-Hubbard Hamiltonian on a three dimensional cubic lattice, and 
compared with the corresponding results from mean-field theory.
Although both approaches yield the same Mott insulator - superfluid phase boundary to first order in the tunneling, 
the predictions of our effective action theory turn out to be superior to the mean-field results deeper into the superfluid 
phase.\end{abstract}
\maketitle
\section{Introduction}
Recent developments in the field of dilute ultracold quantum gasses  \cite{boseold7,contex,boseold8,lewenstein,grimm} have led to 
the experimental investigation of atoms in periodic potentials \cite{mandel}. They are a fascinating new generation of many-particle 
quantum systems as they allow for the study of a variety of solid-state phenomena under perfectly controlled conditions 
\cite{lewenstein,mandel,collapse,gerbierpra,gerbier,FW06,zurich,OOW06}. For instance, Bosonic lattice systems show a quantum phase 
transition for varying lattice depths. In deep lattices the tunneling between lattice sites is suppressed, and a Mott insulating state 
forms with a fixed number of Bosons residing on each lattice site. For shallow lattices, however, the dominance of inter-site tunneling 
allows for Bosons to coherently spread over the whole lattice, forming a superfluid. The occurrence of such a quantum phase transition 
between a Mott insulator and a superfluid is observable, for instance, in time-of-flight absorption pictures taken after switching off 
the lattice potential. They image momentum distributions integrated along one axis, and therefore by Heisenberg's uncertainty principle 
give information about the corresponding spacial distributions. Thus, the localization of atoms in the Mott phase results in diffuse 
absorption pictures, while the delocalized superfluid phase gives rise to Bragg-like interference patterns.

The theoretical analysis of this quantum phase transition is usually based on the Bose-Hubbard model Hamiltonian 
\cite{fischer,jaksch,sachdev,zoller},
\begin{equation}
\hat{H}_\mathrm{BH}=\sum_i{\left[\frac{1}{2}U\hat{a}^{\dag}_i\hat{a}_i\Big(\hat{a}^{\dag}_i\hat{a}_i-1\Big)
-\mu\hat{a}^{\dag}_i\hat{a}_i\right]}-t\sum_{\left<i,j\right>}{\hat{a}^{\dag}_i\hat{a}_j}, \label{eq23}
\end{equation}
where $\hat{a}_i$ and $\hat{a}^{\dag}_i$ are Bosonic annihilation and creation operators,
$\mu$ is the chemical potential,
 and $\left<i, j\right>$ signifies a sum over nearest neighbor sites $i$ and $j$. Additionally, $U$ parameterizes the on-site 
interaction energy between two atoms at a given site, and $t$ characterizes the kinetic energy, in this case given by the tunneling 
of an atom between two neighboring lattice sites.
 The quartic on-site coupling term, however, makes an exact diagonalization of (\ref{eq23}) impossible. 
Thus, while Monte-Carlo simulations have proven fruitful for obtaining numerical results \cite{montecarlo1,montecarlo2,montecarlo3}, 
analytic descriptions of Bosonic lattices near the quantum phase boundary have so far been typically limited to mean-field 
\cite{fischer,sachdev} or strong-coupling approximations \cite{monien,monien2}. Currently, the most precise analytic result for the 
whole Mott insulator-superfluid phase diagram in a three dimensional cubic lattice at zero temperature is found in 
Ref.~\cite{ednilson}. Therein, a Landau expansion for an effective potential with a spatially and temporally global order parameter 
is derived. 
In this paper, we generalize the results of Ref.~\cite{ednilson} by allowing for a spacially and temporally varying order parameter, 
thus determining a Ginzburg-Landau expansion for the effective action. This allows us to obtain an approximate analytic description of 
Bosonic lattice systems near the quantum phase boundary.

To this end we proceed as follows.
In Section \ref{sec1}, we consider a very general type of Hamiltonian consisting of an arbitrary on-site interaction and an arbitrary 
tunneling term, 
of which the Bose-Hubbard Hamiltonian is a special case, and determine the grand-canonical free energy to first order in the tunneling 
term. This tunneling approximation is motivated by the fact that in three dimensions, the Mott insulator-superfluid quantum phase 
transition is observed to occur for small values of ${t}/{U}$ (note that the tunneling expansion is related to the random-walk 
expansion of Refs. \cite{ziegler1,ziegler2,ziegler3}).
Next, in Section \ref{sec2} we introduce an order parameter field and derive a Ginzburg-Landau expansion of the effective action, 
allowing for the computation of physical quantities near the phase boundary in both the Mott insulator and the superfluid phase.
Sections \ref{sec3} and \ref{sec4} present predictions of our effective-action theory for both static homogeneous and 
spatio-temporally varying order parameter fields, including expressions for the particle density, the compressibility, the superfluid 
density, and the excitation spectra. 
Finally, in Section \ref{sec5}, we specify our results to the Bose-Hubbard Hamiltonian, and compare them to the predictions of the 
standard mean-field theory. Although both approaches yield the same approximation for the location of the phase boundary, our effective 
action approach turns out to be superior to the mean-field theory for the following reasons. First, we demonstrate that the 
effective action approach leads to qualitatively better results deeper into the superfluid phase. Secondly, in contrast to the 
mean-field approximation, the effective action approach can be systematically extended to higher orders in the tunneling parameter 
in order to quantitatively improve the results, as has already been demonstrated for the case of the effective potential in 
Ref.~\cite{ednilson}.
\section{Grand-Canonical Free Energy} \label{sec1}
We consider Bosons on a background lattice with lattice sites denoted by $i$. 
Suppose they are described by a Hamiltonian of the form
\begin{equation}
\hat{H}=\hat{H}_0+\hat{H}_1, \label{eq1}
\end{equation}
which depends on Bosonic creation and annihilation operators $\hat{a}^{\dag}_i$ and $\hat{a}_i$ obeying the standard commutation 
relations
\begin{equation}
\Big[\hat{a}_i,\hat{a}_j\Big]=\left[\hat{a}^{\dag}_i,\hat{a}^{\dag}_j\right]=0,\,\,\,\left[\hat{a}_i,\hat{a}^{\dag}_j\right]=\delta_{ij}.
\end{equation}
We assume that $\hat{H}_0$ is a sum of local terms each diagonal in the occupation number basis, i.e. 
\begin{equation}
\hat{H}_0=\sum_i{f_i(\hat{a}^{\dagger}_i\hat{a}_i)} \label{eqx}
\end{equation}
so its energy eigenvalues are given by
\begin{equation}
E_{\{n_i\}}=\sum_{i}f_i(n_i).
\end{equation}
As we will be working grand-canonically, we stipulate that the terms $f_i(\hat{a}^\dag_i\hat{a}_i)$ include the usual 
$-\mu\hat{a}^{\dag}_i\hat{a}_i$ dependence on the chemical potential $\mu$. 
Also, we make the simplifying assumption that $\hat{H}_1$ is only a two-Boson hopping term
\begin{align}
\hat{H}_1 &=-\sum_{ij}{t_{ij}\hat{a}^{\dagger}_i\hat{a}_j} \label{eqa}
\end{align}
with $t_{ij}$ symmetric in $i$ and $j$ and $t_{ii} =0$. 
As with the Bose-Hubbard model (\ref{eq23}), $\hat{H}_0$ in Eq.~(\ref{eqx}) describes the Bosonic on-site interaction, while 
$\hat{H}_1$ in Eq.~(\ref{eqa}) incorporates the tunneling of Bosons between lattice sites. 
Note, however, that Eq.~(\ref{eq1}) with Eqs.~(\ref{eqx}) and (\ref{eqa}) covers a significantly more general scenario than the 
Bose-Hubbard model. 
The on-site interaction in the Bose-Hubbard model (\ref{eq23}) is a two-Boson term with a global interaction strength, but in 
Eq.~(\ref{eqx}), however, we have allowed for the on-site interaction of any finite number of Bosons. 
In addition, we have allowed the Hamiltonian to vary between lattice sites. 
Thus, our model is also capable of describing on-site disorder, which may arise from a local chemical potential, or from a local 
interaction \cite{lewenstein,fischer,krutitsky}. 
Furthermore, Eq.~(\ref{eqa}) contains not only the tunneling of Bosons between nearest neighbor sites as in Eq.~(\ref{eq1}), but 
also between arbitrarily distant sites. 

As we are ultimately interested in investigating quantum phase transitions, we follow general field-theoretic considerations and 
add source terms to the Hamiltonian (\ref{eq1}) in order to explicitly break any global symmetries \cite{kleinert2,zinn-justin}
\begin{align}
\hat{H}_1\rightarrow \hat{H}_1'=-\sum_{ij}{t_{ij}\hat{a}^{\dagger}_i\hat{a}_j}
+\sum_i{\left[j_i(\tau)\hat{a}^{\dagger}_i+j^{*}_i(\tau)\hat{a}_i\right]}. \label{eq2}
\end{align}
Since the source currents $j_i(\tau),j^{*}_i(\tau)$ depend explicitly upon the imaginary time variable $\tau$, standard 
time-dependent perturbation theory may be used to find a perturbative expression for the grand-canonical free energy. 
To do this, we switch to the imaginary-time Dirac interaction picture \cite{peskin}, with operators given by
\begin{equation}
\hat{O}_\mathrm{D}(\tau)=e^{\tau\hat{H}_0}\hat{O}e^{-\tau\hat{H}_0}, \label{eqdirac}
\end{equation}
where we have set $\hbar=1$. In this representation, the Schr\"odinger initial value problem for the time-evolution operator 
takes the form
\begin{align}
\frac{d}{d\tau}\hat{U}_\mathrm{D}(\tau,\tau_0)&=-\hat{H}_{1\mathrm{D}}'(\tau)\hat{U}_\mathrm{D}(\tau,\tau_0), \\
\hat{U}_\mathrm{D}(\tau_0,\tau_0)&=1.
\end{align}
This is solved by the Dyson expansion
\begin{align}
\hat{U}_\mathrm{D}(\tau,\tau_0) & =1+\sum_{n=1}^{\infty}{\hat{U}_\mathrm{D}^{(n)}(\tau,\tau_0)}, \label{eq2pprime} \\
\hat{U}_\mathrm{D}^{(n)}(\tau,\tau_0) & =\frac{(-1)^n}{n!} 
\int_{\tau_0}^{\tau}{\!\!\!\!\mathrm{d}\tau_1}\int_{\tau_0}^{\tau}{\!\!\!\!\mathrm{d}\tau_2}\dots\int_{\tau_0}^{\tau}
{\!\!\!\!\mathrm{d}\tau_n\hat{T}\left[\hat{H}_{1\mathrm{D}}'(\tau_1)\hat{H}_{1\mathrm{D}}'(\tau_2)
\dots\hat{H}_{1\mathrm{D}}'(\tau_n)\right]}, \label{eq2prime}
\end{align}
where $\hat{T}$ is the standard imaginary-time ordering operator. 
The grand-canonical partition function for the system is defined as 
\begin{equation}
\mathcal{Z}=\mathrm{tr}\left\{\hat{T}e^{-\int_{0}^{\beta}\mathrm{d}\tau\hat{H}(\tau)}\right\},
\end{equation}
which can be rewritten as
\begin{equation}
\mathcal{Z}=\mathrm{tr}\left\{e^{-\beta\hat{H}_0}\hat{U}_\mathrm{D}(\beta,0)\right\}. \label{eq3}
\end{equation}
This gives the partition function $\mathcal{Z}$ as a functional of the currents. 
For brevity, we shall - in cases where no confusion may arise - suppress the arguments of functionals. 
Thus, by substituting Eqs.~(\ref{eq2pprime}) and (\ref{eq2prime}) into Eq.~(\ref{eq3}),  we obtain
\begin{align}
\mathcal{Z} & =\mathcal{Z}^{(0)}+\sum_{n=1}^{\infty}{\mathcal{Z}^{(n)}}, \\
\mathcal{Z}^{(n)} & =\mathcal{Z}^{(0)}\frac{(-1)^n}{n!} 
\int_{0}^{\beta}{\!\!\!\!\mathrm{d}\tau_1}\int_{0}^{\beta}{\!\!\!\!\mathrm{d}\tau_2}\dots\int_{0}^{\beta}
{\!\!\!\!\mathrm{d}\tau_n\left<\hat{T}\left[\hat{H}_{1\mathrm{D}}'(\tau_1)
\hat{H}_{1\mathrm{D}}'(\tau_2)\dots\hat{H}_{1\mathrm{D}}'(\tau_n)\right]\right>_0}, \label{eq4}
\end{align}
where 
\begin{equation}
\mathcal{Z}^{(0)}_i=\mathrm{tr}\{e^{-\beta\hat{H}_0}\}=\prod_i
\sum_{n=0}^{\infty}e^{-\beta f_i(n)}
\end{equation}
is the partition function of the unperturbed system, and 
\begin{equation}
<\bullet>_0=\frac{1}{\mathcal{Z}^{(0)}}\mathrm{tr}\left\{\bullet \,e^{-\beta\hat{H}_0}\right\}
\end{equation}
 represents the thermal average with respect to the \emph{unperturbed} Hamiltonian $\hat{H}_0$. 
 This can be expressed more compactly as
\begin{equation}
\mathcal{Z}=\mathcal{Z}^{(0)}\left<\hat{T}\exp\left(-\int_{0}^{\beta}{\!\!\!\!\mathrm{d}\tau\hat{H}'_{1\mathrm{D}}(\tau)}\right)\right>_0.
\end{equation}
Inserting the explicit form of $\hat{H}_{1\mathrm{D}}'(\tau)$ from Eqs.~(\ref{eq2}) and (\ref{eqdirac}), we see that the 
expectation values appearing in Eq.~(\ref{eq4}) can be expanded in terms of Green's functions of the unperturbed system. 
Furthermore, since the grand-canonical free energy is given as a logarithm of the partition function
\begin{equation}
\mathcal{F}=-\frac{1}{\beta}\log\mathcal{Z},
\end{equation}
the Linked Cluster Theorem \cite{gelfand} tells us that $\mathcal{F}$ can be expanded diagrammatically in terms of 
\emph{cumulants} defined as
\begin{align}
C_{2n}^{(0)}(i_1',\tau_1';\dots;i_n',\tau_n'| i_1,\tau_1;\dots;i_n,\tau_n) = \left.\frac{\delta^{2n}C_0^{(0)}
[j,j^*]}{\delta j_{i'_1}(\tau'_1)\dots\delta j_{i'_n}(\tau'_n)\delta j^*_{i_1}(\tau_1)\dots\delta j^*_{i_n}(\tau_n)}\right|_{j=j^*=0}, 
\label{eqb}
\end{align}
with the generating functional
\begin{align}
C_0^{(0)}[j,j^*] = \log \left.\frac{\mathcal{Z}}{\mathcal{Z}^{(0)}}\right|_{t_{ij}\equiv 0}=\log{\left<\hat{T}
\exp{\left\{-\left(\sum_i{\int_0^{\beta}
{\mathrm{d}\tau\left[j_i(\tau)\hat{a}^{\dag}_i(\tau)+j_i^{*}(\tau)\hat{a}_i(\tau)\right]}}\right)\right\}}\right>_0}, \label{eq8}
\end{align}
with only contributions from connected diagrams \cite{metzner}. 
Note that this approach, rather than a decomposition of the Green's functions via Wick's theorem, must be used in our case 
as $\hat{H}_0$ is not necessarily quadratic in the creation and annihilation operators. 
Because $\hat{H}_0$ is local according to Eq.~(\ref{eqx}), the average in Eq.~(\ref{eq8}) factors into independent averages for 
each lattice site. 
It follows that $C_0^{(0)}[j,j^*]$ is a sum of local quantities, and thus the cumulants 
$C_{2n}^{(0)}(i_1',\tau_1';\dots;i_n',\tau_n'| i_1,\tau_1;\dots;i_n,\tau_n)$ vanish unless all site indices are equal. 
With this, we can write
\begin{equation}
C_{2n}^{(0)}(i_1',\tau_1';\dots;i_n',\tau_n'| i_1,\tau_1;\dots;i_n,\tau_n)
={_{i_1}C}_{2n}^{(0)}(\tau_1',\dots,\tau_n'| \tau_1,\dots,\tau_n)\prod_{n,m}{\delta_{i'_n,i_m}},
\end{equation}
so that it only remains to determine the local quantities $_iC_{2n}^{(0)}(\tau_1',\dots,\tau_n'| \tau_1,\dots,\tau_n)$. 
Using the definitions (\ref{eqb}) and (\ref{eq8}), we find that
\begin{equation}
_iC^{(0)}_2(\tau_1|\tau_2)=\left<\hat{T}\left[\hat{a}^{\dag}_i(\tau_1)\hat{a}_i(\tau_2)\right]\right>_0=
G^{(0)}(i,\tau_1|i,\tau_2), \label{eqac}
\end{equation}
where $G^{(0)}(i,\tau_1|j,\tau_2)=\delta_{ij}G^{(0)}(i,\tau_1|i,\tau_2)$ is the imaginary-time Green's function of the unperturbed system. 
Similarly,
\begin{align}
_iC^{(0)}_4(\tau_1,\tau_2|\tau_3,\tau_4)=&\left<\hat{T}\left[\hat{a}^{\dag}_i(\tau_1)\hat{a}^{\dag}_i(\tau_2)\hat{a}_i(\tau_3)
\hat{a}_i(\tau_4)\right]\right>_0 \nonumber \\
&-{_iC}^{(0)}_2(\tau_1|\tau_3)_iC^{(0)}_2(\tau_2|\tau_4)-_iC^{(0)}_2(\tau_1|\tau_4)_iC^{(0)}_2(\tau_2|\tau_3). \label{eqad}
\end{align}
Note that local the quantity $_iC^{(0)}_4(\tau_1,\tau_2|\tau_3,\tau_4)$ is symmetric under both the exchanges 
$\tau_1\leftrightarrow\tau_2$ and $\tau_3\leftrightarrow\tau_4$.

Because each power of the tunneling parameter $t_{ij}$ is associated with a creation operator and an annihilation operator, 
and each power of $j_i(\tau)$ ($j^*_i(\tau)$) is associated with one creation (annihilation) operator, we can construct the 
connected diagrams which contribute to $\mathcal{F}$ according to the following rules \cite{matthias}:
\begin{enumerate}
\item Each vertex with $n$ lines entering and $n$ lines exiting corresponds to a $2n$-th order cumulant $_iC_{2n}^{(0)}$.
\item Draw all topologically inequivalent connected diagrams.
\item Label each vertex with a site index, and each line with an imaginary-time variable.
\item Each internal line is associated with a factor of $t_{ij}$.
\item Each incoming (outgoing) external line is associated with a factor of $j_i(\tau)$ ($j^*_i(\tau)$).
\item Multiply by the  multiplicity and divide by the symmetry factor.
\item Integrate over all internal time variables.
\end{enumerate}
Each diagram is then multiplied by the appropriate factors of $j_i(\tau)$, $j_i^*(\tau)$, and $t_{ij}$, and all spacetime variables 
are integrated. 
Since $\hat{H}_0$ in Eq.~(\ref{eqx}) is diagonal in the occupation number basis and local, there can be no contributions from 
diagrams with one line. 
Thus, to first order in the tunneling $t_{ij}$ and fourth order in the currents $j_i(\tau)$ we find
\begin{align} 
\mathcal{F}=&F_0-\frac{1}{\beta}\sum_i{}\left\{\int_0^{\beta}{\!\!\!\!\mathrm{d}\tau_1}\int_0^{\beta}
{\!\!\!\!\mathrm{d}\tau_2\left[a_{2}^{(0)}(i,\tau_1|i,\tau_2)j_i(\tau_1)j^*_i(\tau_2)+\sum_j{a_{2}^{(1)}(i,\tau_1|j,\tau_2)
t_{ij}j_i(\tau_1)j^*_j(\tau_2)}\right]}\right. \nonumber \\
&+\frac{1}{4}\int_0^{\beta}{\!\!\!\!\mathrm{d}\tau_1}\!\!\int_0^{\beta}{\!\!\!\!\mathrm{d}\tau_2}\!\!\int_0^{\beta}{\!\!\!\!\mathrm{d}
\tau_3}
\!\!\int_0^{\beta}{\!\!\!\!\mathrm{d}\tau_4\,a_4^{(0)}(i,\tau_1;i,\tau_2|i,\tau_3;i,\tau_4)j_i(\tau_1)j_i(\tau_2)j^*_i(\tau_3)j^{*}_i
(\tau_4)} \nonumber \\
&+\frac{1}{2}\left.\int_0^{\beta}{\!\!\!\!\mathrm{d}\tau_1}\!\!\int_0^{\beta}{\!\!\!\!\mathrm{d}\tau_2}\!\!
\int_0^{\beta}{\!\!\!\!\mathrm{d}\tau_3}\!\!\int_0^{\beta}{\!\!\!\!\mathrm{d}\tau_4}\sum_j\, t_{ij}
\left[a_{4}^{(1)}(i,\tau_1;i,\tau_2|j,\tau_3;i,\tau_4)j_i(\tau_1)j_i(\tau_2)j^*_j(\tau_3)j^*_i(\tau_4)\right.\right.\nonumber \\
&+\left.\left.a_{4}^{(1)}(i,\tau_1;j,\tau_2|i,\tau_3;i,\tau_4)j_i(\tau_1)j_j(\tau_2)j^*_i(\tau_3)j^*_i(\tau_4)\right]\right\}, \label{eq5}
\end{align}
where
\begin{equation}
F_0=-\frac{1}{\beta}\log\mathcal{Z}^{(0)}=-\frac{1}{\beta}\sum_i\log\left\{\sum_{n=0}^{\infty}e^{-\beta f_i(n)}\right\} \label{eqf0}
\end{equation}
is the grand-canonical free energy of the unperturbed system, and the respective coefficients $a_{2n}$ are given by the following 
diagrams and expressions:
\begin{align}
a_2^{(0)}(i,\tau_1|i,\tau_2)&=\;\;\;\;\;\begin{fmffile} {a2}
\begin{fmfgraph*}(25,1)
\fmfleft{i1}
\fmfright{o1}
\fmf{fermion}{i1,v1,o1}
\fmfdot{v1}
\fmflabel{$\tau_1$}{i1}
\fmflabel{$\tau_2$}{o1}
\fmfv{label=$i$,label.angle=90}{v1} 
\end{fmfgraph*}
\end{fmffile}\;\;\;\;\;={_iC}_{2}^{(0)}(\tau_1|\tau_2), \label{eqaa} \\
a_2^{(1)}(i,\tau_1|j,\tau_2)&=\;\;\;\;\;\begin{fmffile}{a21}
\begin{fmfgraph*}(25,1)
\fmfleft{i1}
\fmfright{o1}
\fmf{fermion}{i1,v1}
\fmf{fermion}{v1,v2}
\fmf{fermion}{v2,o1}
\fmfdot{v1}
\fmfdot{v2}
\fmflabel{$\tau_1$}{i1}
\fmflabel{$\tau_2$}{o1}
\fmfv{label=$i$,label.angle=90}{v1} 
\fmfv{label=$j$,label.angle=90}{v2} 
\end{fmfgraph*}
\end{fmffile}\;\;\;\;\;=\int_0^{\beta}{\mathrm{d}\tau {_iC}_2^{(0)}(\tau_1|\tau){_jC}_2^{(0)}(\tau|\tau_2)}, \label{eqaaa} \\ 
\nonumber \\
\nonumber \\
a_4^{(0)}(i,\tau_1;i,\tau_2|i,\tau_3;i,\tau_4)&=\;\;\;\;\;\begin{fmffile}{a4}
\begin{fmfgraph*}(25,10)
\fmfleft{i1,i2}
\fmfright{o1,o2}
\fmf{fermion}{i1,v1,o2}
\fmf{fermion}{i2,v2,o1}
\fmf{photon}{v1,v2}
\fmfdot{v1}
\fmflabel{$\tau_1$}{i1}
\fmflabel{$\tau_2$}{i2}
\fmflabel{$\tau_3$}{o2}
\fmflabel{$\tau_4$}{o1}
\fmfv{label=$i$,label.angle=90}{v1} 
\end{fmfgraph*}
\end{fmffile}\;\;\;\;\;={_iC}_4^{(0)}(\tau_1,\tau_2|\tau_3,\tau_4), \label{eqab} \\
\nonumber \\
\nonumber \\
a_4^{(1)}(i,\tau_1;i,\tau_2|j,\tau_3;i,\tau_4)\label{eq6}&=\;\;\;\;\;\begin{fmffile}{a41}
\begin{fmfgraph*}(25,10)
\fmfleft{i1,i2}
\fmfright{o1,o2}
\fmf{fermion,tension=1}{i1,v1}
\fmf{fermion}{v1,v2}
\fmf{fermion}{v2,o2}
\fmf{fermion,tension=1}{i2,v1}
\fmfdot{v1}
\fmfdot{v2}
\fmf{fermion}{v1,o1}
\fmflabel{$\tau_1$}{i1}
\fmflabel{$\tau_2$}{i2}
\fmflabel{$\tau_3$}{o2}
\fmflabel{$\tau_4$}{o1}
\fmfv{label=$i$,label.angle=90}{v1} 
\fmfv{label=$j$,label.angle=90}{v2}
\end{fmfgraph*}
\end{fmffile}\;\;\;\;\;=\int_0^{\beta}{\mathrm{d}\tau{ _iC}_4^{(0)}(\tau_1,\tau_2|\tau,\tau_4){_jC}_2^{(0)}(\tau|\tau_3)}.
\end{align}
In the next section, we will see how this expansion (\ref{eq5}) of the grand-canonical free energy leads to the Ginzburg-Landau 
expansion of the effective action to first order in $t_{ij}$.

At this point, it is also worth making some observations about the two-particle Green's function $G(i,\tau_1|j,\tau_2)$, defined 
in the standard way as
\begin{equation}
G(i,\tau_1|j,\tau_2)=\left<\hat{T}\left[\hat{a}^{\dag}_i(\tau_1)\hat{a}_j(\tau_2)\right]\right>
=-\beta\frac{\delta^2\mathcal{F}}{\delta j_i(\tau_1)\delta j^*_j(\tau_2)}=\frac{1}{\mathcal{Z}}
\mathrm{tr}\left\{e^{-\beta \hat{H}_0}\frac{\delta^2\hat{U}_\mathrm{D}(\beta,0)}{\delta j_i(\tau_1)\delta j^*_j(\tau_2)}\right\}. 
\label{eqc}
\end{equation}
This quantity can also be expanded diagrammatically in terms of cumulants, provided we realize that the effect of the prefactor 
$1/{\mathcal{Z}}$ in Eq.~(\ref{eqc}) is simply to cancel all disconnected diagrams \cite{peskin}, 
ensuring that the only diagrams that contribute are connected diagrams with two external lines \cite{metzner,matthias}. 
Thus, there is a natural correspondence between the Green's function and the coefficients $a_2^{(n)}$ defined above:
\begin{align}
G(i,\tau_1|j,\tau_2) &=G^{(0)}(i,\tau_1|j,\tau_2)+G^{(1)}(i,\tau_1|j,\tau_2)+\dots \nonumber \\
&=\delta_{ij}a_2^{(0)}(i,\tau_1|i,\tau_2)+a_2^{(1)}(i,\tau_1|j,\tau_2)+\dots .
\end{align}
\section{Effective Action} \label{sec2}
Evaluation of the diagrams shown in Eqs.~(\ref{eqaaa}) and (\ref{eq6}) involve integration over the time variable associated with 
the internal line. 
Thus, their evaluation can be simplified by transforming to Matsubara space, where these integrals amount to simple multiplication. 
We use the following convention for the forward and inverse Matsubara transformations
\begin{align}
g(\omega_m) &=\frac{1}{\sqrt{\beta}}\int_0^{\beta}{\!\!\!\!\mathrm{d}\tau\,e^{i\omega_m\tau}g(\tau)}, \label{eq7} \\
g(\tau)&=\frac{1}{\sqrt{\beta}}\sum_{m=-\infty}^{\infty}{g(\omega_m)e^{-i\omega_m\tau}} \label{eq88},
\end{align}
with the Matsubara frequencies
\begin{align}
\omega_m&=\frac{2\pi m}{\beta}\,\,\, , \,\,\, m\in\mathbb{Z}.
\end{align}
Since the unperturbed Hamiltonian (\ref{eqx}) is time-translation invariant, it follows from Eqs.~(\ref{eqb}) and (\ref{eq8}) that 
the cumulants - and thus the functions $a_{2n}$ - must depend on time-differences only. 
In terms of Matsubara frequencies, this implies that  one of the frequency variables is restricted by a delta function, i.e. we have
\begin{align}
a_2^{(0)}(i,\omega_{m1}|i,\omega_{m2})&=a_2^{(0)}(i,\omega_{m1})\delta_{\omega_{m1},\omega_{m2}}, \\
a_4^{(0)} (i,\omega_{m1};i,\omega_{m2}|i,\omega_{m3};i,\omega_{m4})&=a_4^{(0)} (i,\omega_{m1};i,\omega_{m2}|i,\omega_{m4})
\delta_{\omega_{m1}+\omega_{m2},\omega_{m3}+\omega_{m4}}.
\end{align}
Using this frequency conservation, we find from Eq.~(\ref{eqaaa})
\begin{align}
a_2^{(1)}(i,\omega_{m1}|j,\omega_{m2})&=a_2^{(0)}(i,\omega_{m1})a_2^{(0)}(j,\omega_{m1})\delta_{\omega_{m1},\omega_{m2}},
\end{align}
and correspondingly Eq.~(\ref{eq6}) implies
\begin{align}
a_4^{(1)}(i,\omega_{m1};i,\omega_{m2}|j,\omega_{m3};i,\omega_{m4})&=a_4^{(0)}(i,\omega_{m1};i,\omega_{m2}|i,\omega_{m4})
a_2^{(0)}(j,\omega_{m3})\delta_{\omega_{m1}+\omega_{m2},\omega_{m3}+\omega_{m4}}.
\end{align}
Thus,  the first order corrections to the functions $a_{2n}$ can be expressed entirely in terms of $a_{2}^{(0)}$ and the 
corresponding zeroth order terms. 
Using the expressions (\ref{eqaa}) and (\ref{eqab}), along with the definitions (\ref{eqac}) and (\ref{eqad}), we find that these 
two coefficients are explicitly given by
\begin{align}
a_2^{(0)}(i,\omega_m)=&\frac{1}{\mathcal{Z}^{(0)}}\sum_{n=0}^{\infty}e^{-\beta f_i(n)} \nonumber \\
&\times\left[\frac{n+1}{f_i(n+1)-f_i(n)-i\omega_{m}}-\frac{n}{f_i(n)-f_i(n-1)-i\omega_{m}}\right], \label{eq24}
\end{align}
and
\begin{align}
a_4^{(0)}&(i,\omega_{m1};i,\omega_{m2}|i,\omega_{m4})=\frac{1}{\beta\mathcal{Z}^{(0)}}\sum_{n=0}^{\infty}{e^{-\beta f_i(n)}} \nonumber \\
&\times\left\{\frac{n(n-1)}{f_i(n-2)-f_i(n-1)+i\omega_{m4}}\left[\frac{1}{i(\omega_{m4}-\omega_{m2})}
\left(\frac{e^{\beta(f_i(n)-f_i(n-1)+i(\omega_{m4}-\omega_{m1}-\omega_{m2}))}-1}{f_i(n)-f_i(n-1)+i(\omega_{m4}-\omega_{m1}-\omega_{m2})} 
\right.\right.\right.\nonumber \\
&-\left.\left.\frac{e^{\beta(f_i(n)-f_i(n-1)-i\omega_{m1})}-1}{f_i(n)-f_i(n-1)-i\omega_{m1}}\right)\right.
-\frac{1}{f_i(n-1)-f_i(n-2)-i\omega_{m2}} \nonumber \\
&\times\left.\left.\left(\frac{e^{\beta(f_i(n)-f_i(n-2)-i(\omega_{m1}+\omega_{m2}))}-1}{f_i(n)-f_i(n-2)-i(\omega_{m1}+\omega_{m2})}
-\frac{e^{\beta(f_i(n)-f_i(n-1)-i\omega_{m1})}-1}{f_i(n)-f_i(n-1)-i\omega_{m1}}\right)\right]\right. \nonumber \\
&+\frac{n^2}{f_i(n)-f_i(n-1)-i\omega_{m2}}\left[\frac{1}{i(\omega_{m4}-\omega_{m2})}
\left(\frac{e^{\beta(f_i(n)-f_i(n-1)+i(\omega_{m4}-\omega_{m1}-\omega_{m2}))}-1}{f_i(n)-f_i(n-1)+
i(\omega_{m4}-\omega_{m1}-\omega_{m2})} \right.\right. \nonumber \\
&-\left.\left.\frac{e^{\beta(f_i(n)-f_i(n-1)-i\omega_{m1})}-1}{f_i(n)-f_i(n-1)-i\omega_{m1}}\right)\right. \nonumber \\
&-\left.\left.\frac{1}{f_i(n-1)-f_i(n)+i\omega_{m4}}\left(\beta\delta_{\omega_{m1}\omega_{m4}}
-\frac{e^{\beta(f_i(n)-f_i(n-1)-i\omega_{m1})}-1}{f_i(n)-f_i(n-1)-i\omega_{m1}}\right)\right]\right. \nonumber \\
&+\frac{n(n+1)}{f_i(n)-f_i(n-1)-i\omega_{m2}}\left[\frac{1}{f_i(n+1)-f_i(n-1)-i(\omega_{m1}+\omega_{m2})}\right. \nonumber \\
&\times\left.\left(\frac{e^{\beta(f_i(n)-f_i(n-1)+i(\omega_{m4}-\omega_{m1}-\omega_{m2}))}-1}{f_i(n)-f_i(n-1)+
i(\omega_{m4}-\omega_{m1}-\omega_{m2})}-\frac{e^{\beta(f_i(n)-f_i(n+1)+i\omega_{m4})}-1}{f_i(n)-f_i(n+1)+i\omega_{m4}}\right)\right. 
\nonumber \\
&-\left.\left.\frac{1}{f_i(n+1)-f_i(n)-i\omega_{m1}}\left(\beta\delta_{\omega_{m1}\omega_{m4}}-
\frac{e^{\beta(f_i(n)-f_i(n+1)+i\omega_{m4})}-1}{f_i(n)-f_i(n+1)+i\omega_{m4}}\right)\right]\right\}_{\omega_{m1}\leftrightarrow\omega_{m2}} 
\nonumber \\
&-\left\{a_2^{(0)}(i,\omega_{m1}|i,\omega_{m4})a_2^{(0)}(i,\omega_{m2})\right\}_{\omega_{m1}\leftrightarrow\omega_{m2}}, \label{eq25}
\end{align}
where we have introduced the notation $\{\bullet\}_{x\leftrightarrow y}$ to denote a symmetrization in the variables $x$ and $y$.  
Hence the expansion of the grand-canonical free energy (\ref{eq5}) can be compactly rewritten as
\begin{align}
\mathcal{F}=F_0-\frac{1}{\beta}\Big[&\sum_{ij}\sum_{\omega_{m1},\omega_{m2}}M_{ij}(\omega_{m1},\omega_{m2})j_i(\omega_{m1})j^*_j(\omega_{m2})
\Big. \nonumber \\
&+\sum_{ijkl}\sum_{\substack{\omega_{m1},\omega_{m2}\\ \omega_{m3},\omega_{m4}}}\Big.N_{ijkl}(\omega_{m1},\omega_{m2},\omega_{m3},\omega_{m4})
j_i(\omega_{m1})j_j(\omega_{m2})j^*_k(\omega_{m3})j^*_l(\omega_{m4})\Big] ,\label{eqgamma}
\end{align}
where we have introduced the abbreviations
\begin{equation}
M_{ij}(\omega_{m1},\omega_{m2})\equiv \left[a_2^{(0)}(i,\omega_{m1})\delta_{ij}
+a_2^{(0)}(i,\omega_{m1})a_2^{(0)}(j,\omega_{m1})t_{ij}\right]\delta_{\omega_{m1}\omega_{m2}} \label{eqalpha}
\end{equation}
and
\begin{align}
N_{ijkl}(\omega_{m1},\omega_{m2},\omega_{m3},\omega_{m4})\equiv& \frac{\delta_{\omega_{m1}+\omega_{m2},\omega_{m3}+
\omega_{m4}}}{4}a_4^{(0)}(i,\omega_{m1};i,\omega_{m2}|i,\omega_{m4})\Big\{\delta_{ij}\delta_{jk}\delta_{kl}\Big. \nonumber \\
&+\left. 2\delta_{il}\left[t_{ik}a_2^{(0)}(k,\omega_{m3})\delta_{ij}+t_{ij}a_2^{(0)}(j,\omega_{m2})\delta_{ik}\right]\right\}. \label{eqbeta}
\end{align}
Use of the expansion given above is limited by the fact that the currents $j_i(\omega_{m})$ are unphysical quantities. 
Therefore we desire a thermodynamic potential in terms of physically relevant observables. 
To this end, we define an order parameter field $\psi_i(\omega_m)$ in the standard field-theoretic way \cite{kleinert2,zinn-justin} as
\begin{equation}
\psi_i(\omega_m)=\left<\hat{a}_i(\omega_m)\right>=\beta\frac{\delta\mathcal{F}}{\delta j^*_i(\omega_m)}.\label{eq10}
\end{equation}
To first order in the tunneling parameter $t_{ij}$, we find that the order parameter field is given by
\begin{align}
\psi_i(\omega_m)=-&\sum_{p}\sum_{\omega_{m1}}M_{pi}(\omega_{m_1},\omega_{m})j_p(\omega_{m_1}) \nonumber \\
&-2\sum_{pjk}\sum_{\omega_{m1},\omega_{m2},\omega_{m3}}N_{pjki}(\omega_{m1},\omega_{m2},\omega_{m3},\omega_m)j_p(\omega_{m1})
j_j(\omega_{m2})j^*_k(\omega_{m3}). \label{eqpsi}
\end{align}
This finding motivates the performance of a Legendre transformation of $\mathcal{F}$ to obtain the effective action which is 
a functional of the order parameter field:
\begin{equation}
\Gamma[\psi_i(\omega_m),\psi^*_i(\omega_m)]=\mathcal{F}-\frac{1}{\beta}\sum_i\sum_{\omega_m}{\left[\psi_i(\omega_m)j^*_i(\omega_m)
+\psi^*_i(\omega_m)j_i(\omega_m)\right]}.\label{eq11}
\end{equation}
The importance of the functional $\Gamma$ is made clear with the following observation. 
The physical situation of interest is the case when the artificially introduced currents vanish, i.e. when we set $j_i(\omega_m)
\equiv j^*_i(\omega_m)\equiv0$. 
Since $\psi$ and $j^*$ are conjugate variables, we have that
\begin{equation}
j_i(\omega_m)=-\beta\frac{\delta\Gamma}{\delta \psi^*_i(\omega_m)}, \label{eqblahblah}
\end{equation}
and thus this physical situation corresponds to
\begin{equation}
\left.\frac{\delta\Gamma}{\delta \psi^*_i(\omega_m)}\right|_{\psi=\psi_{\mathrm{eq}}}\equiv
\left.\frac{\delta\Gamma}{\delta \psi_i(\omega_m)}\right|_{\psi=\psi_{\mathrm{eq}}}\equiv0. \label{eq13}
\end{equation}
This means that the equilibrium value of the square of the order parameter field $\left|\psi\right|^2_{\mathrm{eq}}$ is determined 
by the condition that the effective action $\Gamma$ is stationary with respect to variations about it. 
Furthermore, we have from Eq.~(\ref{eq11}) that the effective action $\Gamma$, evaluated at the equilibrium order parameter field, 
is equal to the physical grand-canonical free energy:
\begin{equation}
\left.\Gamma\right|_{\psi=\psi_{\mathrm{eq}}}=\lim_{j\rightarrow0}\mathcal{F}. \label{eq50}
\end{equation}
Now, a Ginzburg-Landau expansion of the effective action can be obtained. First, using the fact that to first order in $t_{ij}$
\begin{equation}
M^{-1}_{ij}(\omega_{m1},\omega_{m2})=\frac{\delta_{\omega_{m1}\omega_{m2}}}{a_2^{(0)}(i,\omega_{m1})}\left[\delta_{ij}-a_2^{(0)}(i,\omega_{m1})
t_{ij}\right],
\end{equation}
Eq.~({\ref{eqpsi}) can be inverted recursively to find $j_i(\omega_m)$ as a functional of the order parameter field, yielding
\begin{align}
j_i(\omega_m)=-&\sum_{p,\omega_{m1}}M^{-1}_{ip}(\omega_m,\omega_{m1})\Bigg[\psi_p(\omega_{m1})\Big. \nonumber \\
&\left.-2\sum_{qjk}\sum_{\omega_{m2},\omega_{m3}}N_{qjkp}(\omega_{m1},\omega_{m2},\omega_{m3},\omega_m)
J_q(\omega_{m1})J_j(\omega_{m2})J^*_k(\omega_{m3})\right], \label{eqj}
\end{align}
where we have defined the abbreviation
\begin{equation}
J_i(\omega_m)=-\sum_{p,\omega_{m1}}M^{-1}_{pi}(\omega_{m1},\omega_m)\psi_p(\omega_{m1}).
\end{equation}
Inserting this expression for $j_i(\omega_m)$ into Eq.~(\ref{eq11}) together with the expansion (\ref{eqgamma}), and keeping terms 
only up to first order in the tunneling $t_{ij}$, we find
\begin{align}
&\Gamma=F_0+\frac{1}{\beta}\sum_i{}\Bigg\{\sum_{\omega_m}\Bigg[{\frac{\left|\psi_i(\omega_m)\right|^2}{a_2^{(0)}(i,\omega_m)}-
\sum_j{t_{ij}\psi_i(\omega_m)\psi^*_j(\omega_m)}}\Bigg]\Bigg. \label{eq12} \\
&-\!\!\Bigg.\sum_{\substack{\omega_{m1},\omega_{m2}\\ \omega_{m3},\omega_{m4}}}
{\frac{a_4^{(0)}(i,\omega_{m1};i,\omega_{m2}|i,\omega_{m3};i,\omega_{m4})}{4a_2^{(0)}(i,\omega_{m1})
a_2^{(0)}(i,\omega_{m2})a_2^{(0)}(i,\omega_{m3})a_2^{(0)}(i,\omega_{m4})}
\psi_i(\omega_{m1})\psi_i(\omega_{m2})\psi^*_i(\omega_{m3})\psi^*_i(\omega_{m4})}\Bigg\}. \nonumber
\end{align}
Thus, after performing the Legendre transformation, it turns out that the tunneling parameter $t_{ij}$ appears up to first order only 
in terms which are quadratic in the order parameter field. 
Furthermore, note that this result for the effective action is sufficiently general that it depends on only three quantities of the 
unperturbed system: 
the grand-canonical free energy (\ref{eqf0}) and the Matsubara transform of the zeroth-order coefficients (\ref{eq24}) and (\ref{eq25}). 
Finally, the condition for equilibrium (\ref{eq13}) becomes
\begin{align}
0&=\frac{\psi_i(\omega_m)}{a_2^{(0)}(i,\omega_m)}-\sum_j t_{ij}\psi_j(\omega_m)
\nonumber \\
&-\sum_{\omega_{m1},\omega_{m2},\omega_{m3}}{\frac{a_4^{(0)}(i,\omega_{m1};i,\omega_{m2}|i,\omega_{m3};i,\omega_m)}
{2a_2^{(0)}(i,\omega_{m1})a_2^{(0)}(i,\omega_{m2})a_2^{(0)}(i,\omega_{m3})a_2^{(0)}(i,\omega_{m})}\psi_i(\omega_{m1})\psi_i(\omega_{m2})
\psi^*_i(\omega_{m3})}. \label{eq14}
\end{align}

Due to the complexity introduced by allowing the functions $f_i$ in the Hamiltonian (\ref{eqx}) to be site-dependent, 
and the fact that many interesting physical scenarios can be modeled with a uniform on-site interaction, we restrict our attention 
in the rest of this paper to the homogeneous situation
\begin{equation}
f_i(\hat{a}^{\dag}_i\hat{a}_i)=f(\hat{a}^{\dag}_i\hat{a}_i).
\end{equation}
In this case, the cumulants are no longer on-site quantities, and we may thus drop the site indices in the coefficients $a_{2n}$. 
In the next sections, we examine the physical implications of both a static and a dynamic order parameter field.
\section{Physical Quantities in the Static Case} \label{sec3}
Consider first an order parameter field that is constant in both time and space, i.e. of the form\begin{equation}
\psi_i(\omega_m)=\psi\sqrt{\beta}\delta_{\omega_m,0}.
\end{equation}
With this, the effective action (\ref{eq12}) simplifies to the effective potential
\begin{equation}
\Gamma=N_s\left[\frac{\left|\psi\right|^2}{a_2^{(0)}(0)}-\frac{\beta 
a_4^{(0)}(0,0|0,0)}{4\left[a_2^{(0)}(0)\right]^4}\left|\psi\right|^4\right]-\left|\psi\right|^2\gamma+F_0, \label{eq15}
\end{equation}
where $N_s$ denotes the total number of lattice sites, and $\gamma=\sum_{ij}{t_{ij}}$. 
In the case where 
\begin{equation}
a_4^{(0)}(0,0|0,0)<0 
\end{equation}
we have, according to the standard Landau theory, a phase transition of second order with a phase boundary given by the set of 
system parameters satisfying
\begin{equation}
0=\frac{N_s}{a_2^{(0)}(0)}-\gamma. \label{eq29}
\end{equation} 
As this is the case of most interest, we will assume that such a phase transition exists. 
We also find that Eq.~(\ref{eq14}) takes the simple form
\begin{equation}
0=\psi\left[\frac{N_s}{a_2^{(0)}(0)}-\gamma-\left|\psi\right|^2\frac{\beta N_sa_4^{(0)}(0,0|0,0)}{2(a_2^{(0)}(0))^4}\right],
\end{equation}
from which we see that in the ordered phase the equilibrium value of $\left|\psi\right|^2$, and thus the condensate density, is given by
\begin{equation}
\left|\psi\right|^2_{\mathrm{eq}}=\frac{2(a_2^{(0)}(0))^3\left[N_s-a_2^{(0)}(0)\gamma\right]}{\beta N_s a_4^{(0)}(0,0|0,0)}.
\label{eq16}
\end{equation}
Furthermore, due to Eq.~(\ref{eq50}), other physical quantities follow from evaluating derivatives of $\Gamma$ at $\psi_{\mathrm{eq}}$. 
For instance, the expectation value of the number of particles per lattice site $\left<n\right>=-\frac{1}{N_s}
\frac{\partial\mathcal{F}}{\partial\mu}$ in the ordered phase is given by
\begin{equation}
\left<n\right>=-\left.\frac{1}{N_s}\frac{\partial\Gamma}{\partial\mu}\right|_{\psi=\psi_{\mathrm{eq}}}, \label{eq32}
\end{equation}
and correspondingly, the compressibility $\kappa=\frac{\partial\left<n\right>}{\partial\mu}$ follows from
\begin{equation}
\kappa=-\left.\frac{1}{N_s}\frac{\partial^2\Gamma}{\partial\mu^2}\right|_{\psi=\psi_{\mathrm{eq}}}. \label{eq33}
\end{equation}
In general, any thermodynamic quantity expressible as a function of derivatives of the grand-canonical free energy $\mathcal{F}$ 
can be expressed as the same function of derivatives of $\Gamma$ with respect to the same variables, 
evaluated at $\psi=\psi_{\mathrm{eq}}$.

Relaxing the condition of spacial homogeneity of the order parameter, we are able to determine the superfluid density of the system. 
The superfluid density is defined as the effective fluid density that remains at rest when the entire system is moved at a 
constant velocity \cite{fischer2,roth}.
As is well known in quantum mechanics, such a uniform velocity corresponds to imposing twisted boundary conditions. 
Equivalently, we introduce Peierls phase factors
\begin{equation}
\hat{a}_i\rightarrow\hat{a}_ie^{i\frac{\vec{x}_i}{L}\cdot\vec{\phi}} \label{eqphase}
\end{equation}
in the original Hamiltonian (\ref{eq1}). 
Here $\vec{\phi}$ is related to the velocity of the system according to $\vec{v}=\vec{\phi}/{m^*L}$ where $m^*$ is the effective 
particle mass, $\vec{x}_i$ are the lattice vectors, and $L$ is the extent of the system in the direction of $\vec{v}$. 
Equating the kinetic energy of the superfluid with the free energy difference $\mathcal{F}(\vec{\phi})-\mathcal{F}(\vec{0})$, we 
see that the superfluid density $\rho$ is given by
\begin{equation}
\rho=\lim_{|\vec{\phi}|\rightarrow0}\frac{2m^*L^2}{N_s|\vec{\phi}|^2}\left[\mathcal{F}(\vec{\phi})-\mathcal{F}(\vec{0})\right]. \label{eq51}
\end{equation}
Examining the form of $\hat{H}_0$ and $\hat{H}_1$ in Eqs.~(\ref{eqx}) and (\ref{eqa}), we see that the effect of introducing the 
phase factors in Eq.~(\ref{eqphase}) is simply to redefine the tunneling parameter $t_{ij}$ as
\begin{equation}
t_{ij}(\vec{\phi})=t_{ij}e^{i\frac{\vec{x}_j-\vec{x}_i}{L}\cdot\vec{\phi}}. \label{eq52}
\end{equation}
Thus, using Eq.~(\ref{eq50}) we can express $\rho$ in terms of the effective action as
\begin{equation}
\rho=\lim_{|\vec{\phi}|\rightarrow0}\frac{2m^*L^2}{N_s|\vec{\phi}|^2}\left[\left.\Gamma(\vec{\phi})\right|_{\psi=\psi_\mathrm{eq}(\vec{\phi})}
-\left.\Gamma(\vec{0})\right|_{\psi=\psi_\mathrm{eq}(\vec{0})}\right], \label{eq53}
\end{equation}
which, with the aid of Eq.~(\ref{eq12}) reduces to
\begin{align}
\rho=&\lim_{|\vec{\phi}|\rightarrow0}\frac{2m^*L^2}{N_s|\vec{\phi}|^2}\left\{\sum_{ij}
\left[t_{ij}\left(\left|\psi_\mathrm{eq}(\vec{\phi})\right|^2e^{i\frac{\vec{x}_j-\vec{x}_i}{L}\cdot\vec{\phi}}
-\left|\psi_\mathrm{eq}(\vec{0})\right|^2\right)\right]\right. \label{eq53a} \\
&+\left.N_s\left[\frac{1}{a_2^{(0)}(0)}\left(\left|\psi_\mathrm{eq}(\vec{\phi})\right|^2-\left|\psi_\mathrm{eq}(\vec{0})\right|^2\right)
-\frac{\beta a_4^{(0)}(0,0|0,0)}{4(a_2^{(0)}(0))^4}\left(\left|\psi_\mathrm{eq}(\vec{\phi})\right|^4
-\left|\psi_\mathrm{eq}(\vec{0})\right|^4\right)\right]\right\} \nonumber.
\end{align}
Thus, the superfluid density is determined explicitly once a definite form of the tunneling parameter $t_{ij}$ is specified.
\section{Physical Quantities in the Dynamic Case} \label{sec4}

By allowing the order parameter to vary in imaginary time, we can also use the effective action to obtain an analytic form for 
the Matsubara Green's function. 
To do so, we note that the Legendre transformation (\ref{eq10}), (\ref{eq11}), (\ref{eqblahblah}) implies
\begin{equation}
{\beta}\left.\frac{\delta^2\Gamma}{\delta\psi_i(\omega_{m1})\delta\psi^*_j(\omega_{m2})}\right|_{\psi=\psi_{\mathrm{eq}}}=
-\left.\frac{\delta j(\psi_j(\omega_{m2}))}{\delta\psi_i(\omega_{m1})}\right|_{\psi=\psi_{\mathrm{eq}}}=
\left.\left(-\beta\frac{\delta^2\mathcal{F}}{\delta j^*_i(\omega_{m1})\delta j_j(\omega_{m2})}\right)^{-1}\right|_{j\equiv0}. \label{eq20}
\end{equation}
We recognize immediately from Eq.~(\ref{eqc}) that this is precisely the inverse of the Matsubara Green's function 
$\mathcal{G}(i,\omega_{m1}|j,\omega_{m2})$. Next, we consider $\Gamma$ expanded to arbitrary order in $t_{ij}$,
\begin{equation}
\Gamma=F_0+\frac{1}{\beta}\sum_i{}\left(\sum_{\omega_m}{\frac{\left|\psi_i(\omega_m)\right|^2}{a_2^{(0)}(\omega_m)}
+\sum_{n=1}^{\infty}\sum_j{\alpha_2^{(n)}(\omega_m)}[(t)^n]_{ij}\psi_i(\omega_m)\psi^*_j(\omega_m)}+\dots \right),
\end{equation}
where the expansion coefficients $\alpha_2^{(n)}$ are determined by methods like those described above. 
We then find from Eq.~(\ref{eq20}) that the Matsubara Green's function is given by
\begin{equation}
\left[\mathcal{G}(j,\omega_{m2}|i,\omega_{m1})\right]^{-1}=
\delta_{\omega_{m1},\omega_{m2}}\left(\frac{\delta_{ij}}{a_2^{(0)}(\omega_{m1})}+\sum_{n=1}^{\infty}{\alpha_2^{(n)}(\omega_{m1})[(t)^n]_{ij}}
+\dots\right). \label{eq35}
\end{equation}
Recognizing that $\delta_{\omega_{m1},\omega_{m2}}\delta_{ij}/{a_2^{(0)}(\omega_{m1})}$ is simply the inverse of the unperturbed Matsubara 
Green's function, 
we see that the power series in $t_{ij}$ in Eq.~(\ref{eq35}) gives a series expansion of the self-energy $\Sigma$:
\begin{equation}
\Sigma(i,\omega_{m1}|j,\omega_{m2})=\left[\mathcal{G}^{(0)}(i,\omega_{m1}|j,\omega_{m2})\right]^{-1}
-\left[\mathcal{G}(i,\omega_{m1}|j,\omega_{m2})\right]^{-1}
=-\delta_{\omega_{m1},\omega_{m2}}\sum_{n=1}^{\infty}{\alpha_2^{(n)}(\omega_{m1})[(t)^n]_{ij}}.
\end{equation}
Thus, we conclude that our effective action gives an expansion for the Green's function in terms of the self-energy 
in powers of $t_{ij}$. 
In the non-ordered phase, this is the same as if we had computed the corrections to the unperturbed Green's function directly 
from our perturbative expansion of $\mathcal{F}$ and performed a resummation \cite{matthias}.  
Specifying Eq.~(\ref{eq35}) to our present first-order case, we hence find
\begin{align}
&\left[\mathcal{G}^{(0)}(j,\omega_{m2}|i,\omega_{m1})\right]^{-1}-\Sigma^{(1)}(j,\omega_{m2}|i,\omega_{m1})  \nonumber \\
&=\delta_{ij}\left[\frac{1}{a_2^{(0)}(\omega_{m1})}+\frac{2\delta_{\omega_{m1}\omega_{m2}}
a_4^{(0)}(\omega_{m1},0|0,\omega_{m2})(a_2^{(0)}(0))^2\left(\frac{\gamma}{N_s}
-\frac{1}{a_2^{(0)}(0)}\right)}{(a_2^{(0)}(\omega_{m1}))^2a_4^{(0)}(0,0|0,0)}\right]-t_{ij}. \label{eq22}
\end{align}
Denoting by $t_{\vec{k}\vec{k}'}$ the Fourier transform of the tunneling parameter,
\begin{equation}
t_{\vec{k}\vec{k}'}=\sum_{ij}t_{ij}e^{i\left(\vec{k}'\cdot\vec{x}_j-\vec{k}\cdot\vec{x}_i\right)},
\end{equation} 
we find that Eq.~(\ref{eq22}) can be rewritten in Fourier space as
\begin{align}
&\left[\mathcal{G}(\vec{k}',\omega_{m2}|\vec{k},\omega_{m1})\right]^{-1}  \nonumber \\
&=\frac{1}{a_2^{(0)}(\omega_{m1})}+\frac{2\delta_{\omega_{m1}\omega_{m2}}a_4^{(0)}(\omega_{m1},0|0,\omega_{m2})(a_2^{(0)}(0))^2
\left(\frac{\gamma}{N_s}
-\frac{1}{a_2^{(0)}(0)}\right)}{(a_2^{(0)}(\omega_{m1}))^2a_4^{(0)}(0,0|0,0)}-t_{\vec{k}\vec{k}'}. \label{eq22f}
\end{align}
We see that the second term above is a contribution to the Green's function due completely to the existence of a non-vanishing 
order parameter. 
This correction can thus be exploited to improve analytical time-of-flight calculations for Bosonic lattice systems in the 
superfluid phase \cite{hoffmann}.

Next, we can examine excitations of the system at zero temperature by looking for spatio-temporal variations of the order parameter 
field about 
$\psi_{\mathrm{eq}}=\sqrt{\left|\psi\right|^2_{\mathrm{eq}}}e^{i\theta_0}$ which preserve the equilibrium condition (\ref{eq13}), 
where $\theta_0$ is an arbitrary global phase. 
To this end, we first must specify to the case where our system is translationally invariant, such that
\begin{equation}
t_{\vec{k}\vec{k'}}=t_{\vec{k}}\delta_{\vec{k}\vec{k'}}.
\end{equation}
Next, we add to the equilibrium value of the order parameter field a small variation $\delta\psi(\vec{x}_i,\omega_m)$. 
We then Taylor expand the effective action $\Gamma$ about $\psi_{\mathrm{eq}}$ in terms of these variations. 
The first order term vanishes due to the equilibrium condition (\ref{eq13}), leaving
\begin{equation}
\Gamma=\Gamma[\psi_{\mathrm{eq}}]+\sum_{\substack{i,j\\ \omega_{m1},\omega_{m2}}}{\left.\frac{\delta^2\Gamma}{\delta(\delta\psi_i(\omega_{m1}))
\delta(\delta\psi^*_j(\omega_{m2}))}\right|_{\delta\psi\equiv0} \delta\psi_i(\omega_{m1})\delta\psi^*_j(\omega_{m2})}+\dots. \label{eqz}
\end{equation}
Demanding that in equilibrium the effective potential is stationary with respect to the variations $\delta\psi$ in the standard way 
gives the equation of motion
\begin{equation}
\sum_{i,\omega_{m1}}{\left.\frac{\delta^2\Gamma}{\delta(\delta\psi_i(\omega_{m1}))
\delta(\delta\psi^*_j(\omega_{m2}))}\right|_{\delta\psi\equiv0} \delta\psi_i(\omega_{m1})}=0.
\end{equation}
This equation can be satisfied in two distinct ways. 
The trivial solution $\delta\psi(\vec{x}_i,\omega_m)\equiv0$ corresponds to the static homogeneous equilibrium examined in the 
previous section. 
The second solution is given by
\begin{equation}
\left.\frac{\delta^2\Gamma}{\delta(\delta\psi_i(\omega_{m1}))\delta(\delta\psi^*_j(\omega_{m2}))}\right|_{\delta\psi\equiv0}=0 \label{eq17}
\end{equation}
and describes the excitation spectrum of the system. 
In particular, by analytically continuing Eq.~(\ref{eq17}) to real frequencies and transforming to Fourier space, 
we are able to identify the dispersion relation of low-lying excitations as those curves $\omega(\vec{k})$ which make the equation 
valid. 
The standard method of performing this analytic continuation is to find the equations of motion in imaginary-time and perform an 
inverse Wick rotation. Because of the complexity of the coefficient $a_4^{(0)}(\omega_{m1},\omega_{m2}|\omega_{m3},\omega_{m4})$, 
however, this is ill-suited to our present needs. Therefore, we note that our imaginary time evolution 
operator $\exp{\left(-\hat{H}\tau\right)}$ can be mapped to the real-time evolution operator $\exp{\left(-i\hat{H}t\right)}$ by 
the formal substitution $\hat{H}\rightarrow i\hat{H}$. 
To maintain the reality of the grand-canonical free energy, we must also perform the 
substitution $\mathcal{F}\rightarrow -i\mathcal{F}$.  
We thus find that in terms of real frequencies the effective action is given by
\begin{align}
&\Gamma_\mathrm{R}=F_0+\frac{1}{\beta}\sum_i{}\left\{\int\!\!\!\mathrm{d}
\omega{\left[-i\frac{\left|\psi_i(\omega)\right|^2}{a_{2\mathrm{R}}^{(0)}(\omega)}-\sum_j{t_{ij}\psi_i(\omega)\psi^*_j(\omega)}\right]}
\right. \label{eqrtef} \\
&+i\!\!\left.\int\!\!\!\mathrm{d}\omega_1\!\!\int\!\!\!
\mathrm{d}\omega_2\!\!\int\!\!\!\mathrm{d}\omega_3\!\!\int\!\!\!\mathrm{d}\omega_4
{\frac{a_{4\mathrm{R}}^{(0)}(\omega_{1},\omega_{2}|,\omega_{3},\omega_{4})}{4a_{2\mathrm{R}}^{(0)}(\omega_{1})
a_{2\mathrm{R}}^{(0)}(\omega_{2})a_{2\mathrm{R}}^{(0)}(\omega_{3})a_{2\mathrm{R}}^{(0)}(\omega_{4})}\psi_i(\omega_{1})
\psi_i(\omega_{2})\psi^*_i(\omega_{3})\psi^*_i(\omega_{4})}\!\!\right\}\!\!, \nonumber
\end{align}
where $a_{2\mathrm{R}}^{(0)}$ and $a_{4\mathrm{R}}^{(0)}$ are obtained from Eqs.~(\ref{eq24}) and (\ref{eq25}) respectively by the 
replacement $f_i(n)\rightarrow if_i(n)$. 
Thus, the real-time continuation of the condition (\ref{eq17}) is given by
\begin{equation}
\left.\frac{\delta^2\Gamma_{\mathrm{R}}}{\delta(\delta\psi_i(\omega_{1}))\delta(\delta\psi^*_j(\omega_{2}))}
\right|_{\delta\psi\equiv0}=0. \label{eq17r}
\end{equation}
In general, the function $\omega(\vec{k})$ will have a positive and a negative frequency branch. Because we determined these 
curves by expanding the effective action $\Gamma_\mathrm{R}$ about a minimum, however,
only the positive frequency branch of $\omega(\vec{k})$ are to be considered as physically relevant.

Since the order parameter is complex, we examine separately variations of both the magnitude and of the phase. 
First, we consider excitations in the amplitude of the order parameter. 
To this end, we replace $\psi$ in Eq.~(\ref{eq12}) by $\psi_{\mathrm{eq}}\sqrt{\beta}\delta_{\omega_m,0}+\delta\psi_i(\omega_m)$, 
where $\delta\psi_i(\omega_m)$ is an arbitrary infinitesimal function of the lattice site $i$ and $\omega_m$, with fixed phase 
$\theta_0$. 
Carrying out the functional derivative in Eq.~(\ref{eq17r}) and performing the continuation outlined above and transforming to 
Fourier space yields the equation
\begin{equation}
0=\frac{-i}{a_{2\mathrm{R}}^{(0)}(\omega_\mathrm{A})}+
\frac{2a_{4\mathrm{R}}^{(0)}(\omega_\mathrm{A},0|0,\omega_\mathrm{A})(a_{2\mathrm{R}}^{(0)}(0))^2\left[\frac{\gamma}{N_s}
+\frac{i}{a_{2\mathrm{R}}^{(0)}(0)}\right]}{(a_{2\mathrm{R}}^{(0)}(\omega_\mathrm{A}))^2a_{4\mathrm{R}}^{(0)}(0,0|0,0)}-t_{\vec{k}}. 
\label{eq18}
\end{equation}
This gives a constraint equation which can be solved for the dispersion relation of amplitude excitations $\omega_\mathrm{A}(\vec{k})$. 
By comparing Eq.~(\ref{eq18}) with the Matsubara Green's function (\ref{eq22f}), 
we notice that the dispersion relation $\omega_A(\vec{k})$ coincides with the poles of the translationally invariant real-time Green's 
function.

To treat the phase degree of freedom, we note first that adding a small time-varying phase to $\psi_{\mathrm{eq}}$ amounts to the 
transformation
\begin{equation}  
\psi\rightarrow\psi_{\mathrm{eq}}e^{i\theta_i(\tau)}\approx\psi_\mathrm{eq}\left[1+i\theta_i(\tau)-\frac{1}{2}\theta_i(\tau)^2\right].
\end{equation}
Expressing this in Matsubara space, we have
\begin{equation}
\psi\rightarrow\psi_\mathrm{eq}\left[1+i\theta_i(\omega_m)-\frac{1}{2}\sum_{\omega_n}\theta_i(\omega_n)
\theta_i(\omega_m-\omega_n)\right]. \label{thetatrans}
\end{equation}
Inserting the transformation (\ref{thetatrans}) into the real-time effective action (\ref{eqrtef}) and performing the 
derivative (\ref{eq17r}) yields the condition
\begin{align}
0=&\frac{-i}{a_{2\mathrm{R}}^{(0)}(\omega_\theta)}+\frac{i}{a_{2\mathrm{R}}^{(0)}(0)}-\frac{2a_{2\mathrm{R}}^{(0)}(0)^4
\left(\frac{\gamma}{N_s}+\frac{i}{a_{2\mathrm{R}}^{(0)}(0)}\right)}{a_{4\mathrm{R}}^{(0)}(0,0|0,0)}\Bigg[2b(0,0,0,0)
+b(\omega_\theta,-\omega_\theta,0,0)\Bigg.\nonumber \\
&\Bigg.+b(0,0,\omega_\theta,-\omega_\theta)-2b(\omega_\theta,0,\omega_\theta,0)-2b(\omega_\theta,0,0,\omega_\theta)\Bigg]
+\frac{\gamma}{N_s}-t_{\vec{k}},\label{eq19}
\end{align}
where we have defined
\begin{equation}
b(\omega_{1},\omega_{2}|\omega_{3},\omega_{4})=
\frac{a_{4\mathrm{R}}^{(0)}(\omega_{1},\omega_{2}|\omega_{3},\omega_{4})}{a_{2\mathrm{R}}^{(0)}(\omega_{1})
a_{2\mathrm{R}}^{(0)}(\omega_{2})a_{2\mathrm{R}}^{(0)}(\omega_{3})a_{2\mathrm{R}}^{(0)}(\omega_{4})}.
\end{equation}
This determines the dispersion relation $\omega_\theta(\vec{k})$ of the phase excitations. 
We note that since $t_{\vec{0}}={\gamma}/{N_s}$, $\omega_\theta(\vec{0})=0$ is a solution to the constraint (\ref{eq19}) 
in accordance with Goldstone's theorem \cite{kleinert2,zinn-justin}.

Lastly, we investigate the phenomenon of second sound. 
As is well known, the observed elementary excitations of a superfluid are given by phonons. 
To obtain their corresponding dispersion relation $\omega_s(\vec{k})$, we must examine the phase excitations in the 
presence of the amplitude variations, i.e. $\psi\rightarrow\left[\psi_\mathrm{eq}+\delta\psi_i(\omega_m)\right]e^{i\theta_i(\omega_m)}$. 
This has been considered, for example, in Refs. \cite{kleinert,weichman}, leading to the result
\begin{equation}
\omega_s(\vec{k})=\sqrt{\omega_A(\vec{k})\omega_\theta(\vec{k})}. \label{eqss}
\end{equation}
\section{An Application: The Bose-Hubbard Hamiltonian} \label{sec5}
Having developed the field-theoretic approach for the general Hamiltonian (\ref{eq1}), (\ref{eqx}), (\ref{eqa}) in the previous 
sections, we are now in a position to apply it to the specific case of the Bose-Hubbard Hamiltonian on a 3-dimensional cubic 
lattice defined 
by Eq.~(\ref{eq23}). As is well known, this model exhibits a quantum phase transition between a Mott insulating phase and a 
superfluid phase \cite{fischer,jaksch,sachdev,zoller}. 
The Hamiltonian (\ref{eq23}) has exactly the form assumed in Section \ref{sec1} when the following identifications are made:
\begin{align}
f(n)&=E_n=\frac{1}{2}Un(n-1)-\mu n \\
t_{ij}&=t \sum_{\sigma}  \left(\delta_{\vec{x}_i,\vec{x}_j+\vec{d}_{\sigma}}
+\delta_{\vec{x}_i,\vec{x}_j-\vec{d}_{\sigma}}\right),
\end{align}
where $\mbox{\boldmath$\sigma$}_k$ denotes the lattice basis vectors with $k=1,2,3$. Additionally, 
we see that the quantity $\gamma$ introduced above simplifies to
\begin{equation}
\gamma=\sum_{ij}{t_{ij}}=6N_st.
\end{equation}
Thus, we can apply all previously derived formulas to extract physical information about the Bose-Hubbard system. 
In the $\omega_m\rightarrow0$ limit, we find that $a_2^{(0)}(0)$ becomes
\begin{equation}
a_2^{(0)}(0)=\frac{1}{\mathcal{Z}^{(0)}}\sum_{n=0}^{\infty}{e^{-\beta E_n}\left(\frac{n+1}{E_{n+1}-E_n}-\frac{n}{E_n-E_{n-1}}\right)} 
\label{eq26}.
\end{equation}
The expression for $a_4^{(0)}(0,0|0,0)$ must be calculated via a more careful limiting procedure. Making use of the limit
\begin{equation}
\lim_{x\rightarrow0}{\frac{e^{bx}-1}{x}}=b,
\end{equation}
we find that
\begin{align}
a_4^{(0)}&(0,0|0,0)=\frac{2}{\beta\mathcal{Z}^{(0)}}\sum_{n=0}^{\infty}{}e^{-\beta E_n}\left\{n(n-1)
\frac{-2}{(E_n-E_{n-1})^2(E_{n}-E_{n-2})}\right. \nonumber \\
&+n^2\left[\frac{2}{(E_n-E_{n-1})^3}+\frac{\beta}{(E_n-E_{n-1})}\right]-n(n+1)\left[\frac{2(E_{n+1}-2E_n+E_{n-1})}{(E_n-E_{n-1})^2
(E_n-E_{n+1})^2}\right. \nonumber \\
&\left.+\frac{2\beta}{(E_{n+1}-E_n)(E_n-E_{n-1})}\right]-(n+1)^2\left[\frac{2}{(E_{n+1}-E_n)^3}-\frac{\beta}{(E_{n+1}-E_n)^2}\right] 
\nonumber \\
&+\left.(n+1)(n+2)\left[\frac{2}{(E_{n+1}-E_n)^2(E_{n+2}-E_n)}\right]\right\}-2(a_2^{(0)}(0))^2. \label{eq27}
\end{align}

\subsection{Effective Action Predictions in the Static Case}
Before examining the physical implications of our effective action approach, we first introduce the standard mean-field treatment 
of the Bose-Hubbard Hamiltonian for comparison. 
The mean-field Hamiltonian is found by performing a Hartree-Fock expansion of the hopping term in the Hamiltonian (\ref{eq23}) 
\cite{fischer,sachdev}. 
Keeping in mind that the order parameter is defined according to $\psi=\left<\hat{a}_i\right>$, this yields
\begin{equation}
\hat{H}_\mathrm{MF}=\hat{H}_0-6t\sum_i{}\left(\psi\hat{a}^{\dag}_i+\psi^*\hat{a}_i-\left|\psi\right|^2\right).\end{equation}
The methods of Section \ref{sec1} can be adapted to give an expansion of the mean-field free energy in powers of the order parameter, 
since
\begin{equation}
\hat{H}_\mathrm{MF}=\hat{H}_0+\hat{H}_1'|_{t=0}+6tN_s\left|\psi\right|^2, \label{eq28}
\end{equation}
when we make the formal identification $j_i(\tau)=-6t\psi$. 
Thus, an expansion of $\mathcal{F}$ to zeroth order in $t$ gives an expansion of the mean-field free energy $\mathcal{F}_{\mathrm{MF}}$ 
in powers of the order parameter, 
provided we recognize that the constant term in Eq.~(\ref{eq28}) contributes a term of order $\left|\psi\right|^2$ 
to $\mathcal{F}_{\mathrm{MF}}$. 
With these considerations in mind, we find the explicit result
\begin{equation}
\mathcal{F}_\mathrm{MF}=F_0-N_s\left(a_2^{\mathrm{MF}}\left|\psi\right|^2+\frac{\beta}{4}a_4^{\mathrm{MF}}\left|\psi\right|^4\right),
\end{equation}
where the mean-field Landau coefficients $a_2^\mathrm{MF}$ and $a_4^\mathrm{MF}$ are given by
\begin{align}
a_2^\mathrm{MF} & = a_2^{(0)}(0)(6t)^2-6t, \\
a_4^\mathrm{MF} & = a_4^{(0)}(0,0|0,0)(6t)^4.
\end{align}
\begin{center}
\begin{figure}
\includegraphics[width=6cm]{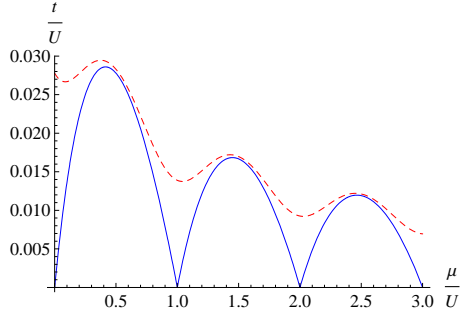}
\caption{Plot of the critical value of the hopping parameter $t$ versus the chemical potential $\mu$, both scaled by the interaction 
energy $U$. 
The phase boundaries for two different temperatures are shown. The solid blue curve is ${T}/{U}=0$, and the dashed red curve is 
${T}/{U}={0.1}/{k_B}$.}\label{fig1}
\end{figure}
\end{center}
Thus, the mean-field result can also be expressed in terms of the same three quantities (\ref{eqf0}),  (\ref{eq24}), and (\ref{eq25}) 
as our effective action approach. 

We now compare the predictions of our mean field theory with our effective action theory. 
First, we find that the mean-field phase boundary is given by the curve \cite{fischer,krutitsky,bru,bousante}
\begin{equation}
t_c^{\mathrm{MF}}=\frac{1}{6a_2^{(0)}(0)}=\frac{\mathcal{Z}^{(0)}}{6\displaystyle\sum_{n=0}^{\infty}{e^{-\beta E_n}
\left(\frac{n+1}{E_{n+1}-E_n}-\frac{n}{E_n-E_{n-1}}\right)}},
\end{equation}
which turns out to be identical to the phase boundary found from Eq.~(\ref{eq29}).  A plot of the phase boundary in Fig.~\ref{fig1} 
reveals that increasing thermal fluctuations destroy quantum coherence, as the superfluid phase shrinks with increasing temperature. 
Note that the main advantage of the field-theoretic method over the mean-field approach is in the fact that the phase boundary can 
be improved by carrying the expansion out to higher orders in $t$. The phase boundary to  second tunneling order has already been 
calculated for $T=0$ in Ref.~\cite{ednilson} and for $T>0$ in Ref.~\cite{matthias}, and proves to be a considerable improvement 
over the mean-field result.

We next look now at the condensate density $\left|\psi\right|^2_{\mathrm{eq}}$. From Eq.~(\ref{eq16}), we  find
\begin{equation}
\left|\psi\right|^2_{\mathrm{eq}}=\frac{2(a_2^{(0)}(0))^3}{\beta a_4^{(0)}(0)}\left[1-6ta_2^{(0)}(0)\right], \label{eq30}
\end{equation}
while standard Landau-theory yields
\begin{equation}
\left|\psi\right|^2_\mathrm{MF}=\frac{-2a_2^{\mathrm{MF}}}{\beta a_4^{\mathrm{MF}}}=\frac{2}{(6t)^3\beta a_4^{(0)}(0)}
\left[1-6ta_2^{(0)}(0)\right] \label{eq31}
\end{equation}
via the minimization of $\mathcal{F}_{\mathrm{MF}}$. 

Turning to the superfluid density, we see from Eq.~(\ref{eq52}) that for the Bose-Hubbard model,
\begin{equation}
t(\vec{\phi})=2t\sum_\sigma{\cos\left(\frac{\vec{d}_\sigma\cdot\vec{\phi}}{L}\right)},
\end{equation}
where $\vec{d}_\sigma$ are the nearest neighbor lattice vectors in the $\sigma$ direction. Therefore, in the Bose-Hubbard model we 
have $t=1/(2m^*)$ \cite{krutitsky2}. Thus, from Eqs.~(\ref{eq51}) and (\ref{eq53a}) we find that for both the mean-field theory and 
the effective action theory to first order in $t$,
\begin{equation}
\rho=\frac{\left|\psi_\mathrm{eq}\right|^2}{|\vec{\phi}|^2}\sum_\sigma{(\hat{x}_\sigma\cdot\vec{\phi})^2}.
\end{equation}
Taking $\vec{\phi}$ to be in a lattice direction, we see that the superfluid and condensate densities are equal at this level of 
approximation.

A plot of the condensate/superfluid density as a function of the tunneling parameter $t$ at a fixed value of the chemical 
potential $\mu$ for each theory can be seen in Fig.~\ref{fig2}.
\begin{figure}
\begin{center}
\includegraphics[width=6cm]{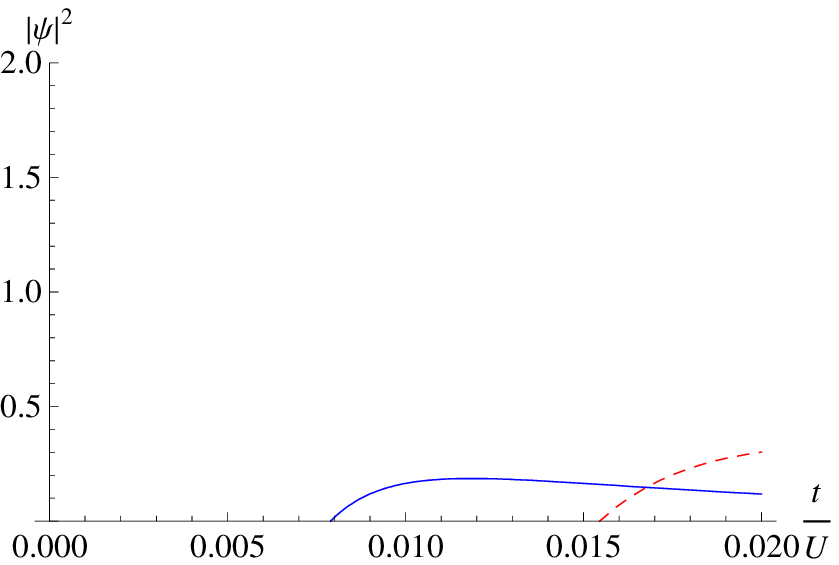}
\hspace{.5cm}
\includegraphics[width=6cm]{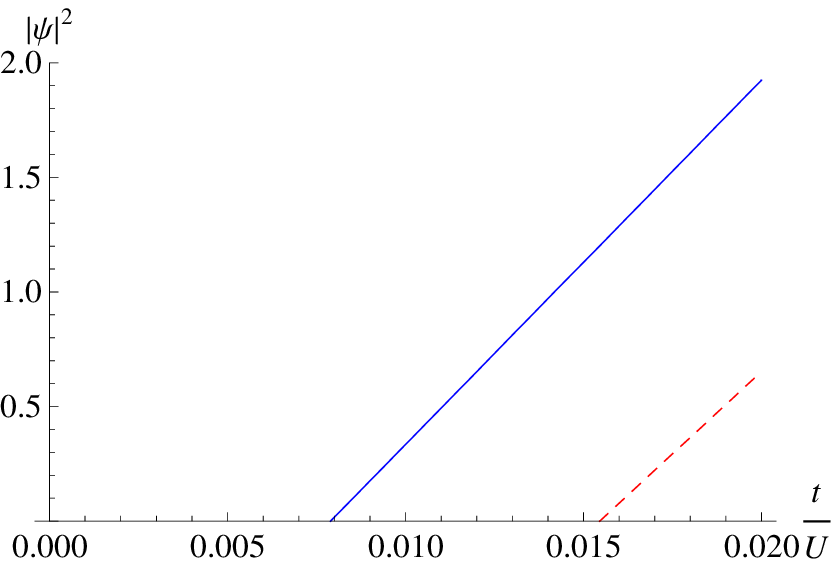}
\caption{Plot of the condensate density as a function of the hopping parameter ${t}/ {U}$ for both the mean-field theory (left) 
and the effective action theory (right), with fixed ${\mu}/ {U}=0.9$. 
The solid blue curves are ${T}/ {U}=0$, and the dashed red curves are ${T}/ {U}={0.1}/ {k_B}$.}\ \label{fig2}
\end{center}
\end{figure}
It is interesting to note that while the superfluid density from the field-theoretic approach increases linearly with $t$, the
 mean-field superfluid density quickly begins to fall off as $t$ increases, a behavior which is at odds with the notion 
of a superfluid \cite{fischer}. 
Furthermore, it can be seen that Eq.~(\ref{eq30}) is simply a first order series expansion of Eq.~(\ref{eq31}) about $t=t_c$, 
meaning  $\left|\psi\right|^2_{\mathrm{eq}}$ is the tangent line to $\left|\psi\right|^2_\mathrm{MF}$ at $t=t_c$. 
Thus, although the two results agree near the phase boundary, the mean-field prediction quickly begins to exhibit an unphysical 
behavior, suggesting that our field-theoretic result has a larger range of validity.

\begin{figure}[h!]
\begin{center}
\includegraphics[width=6cm]{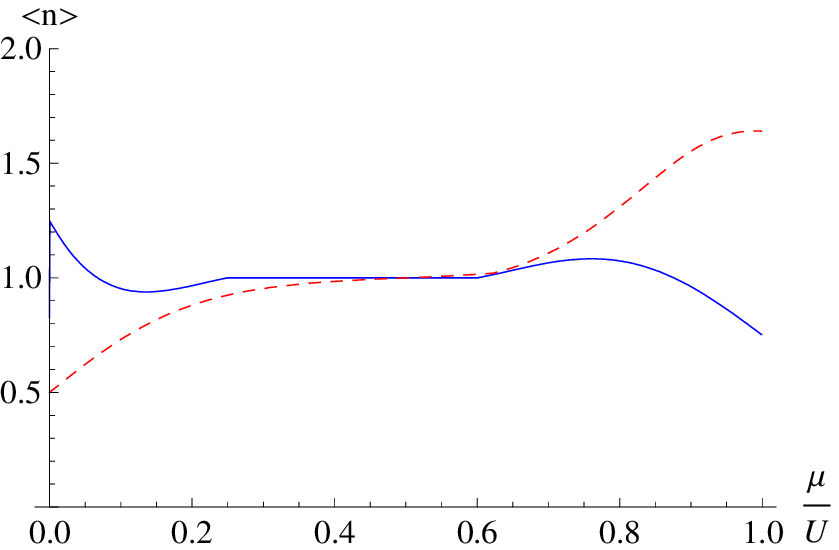}
\includegraphics[width=6cm]{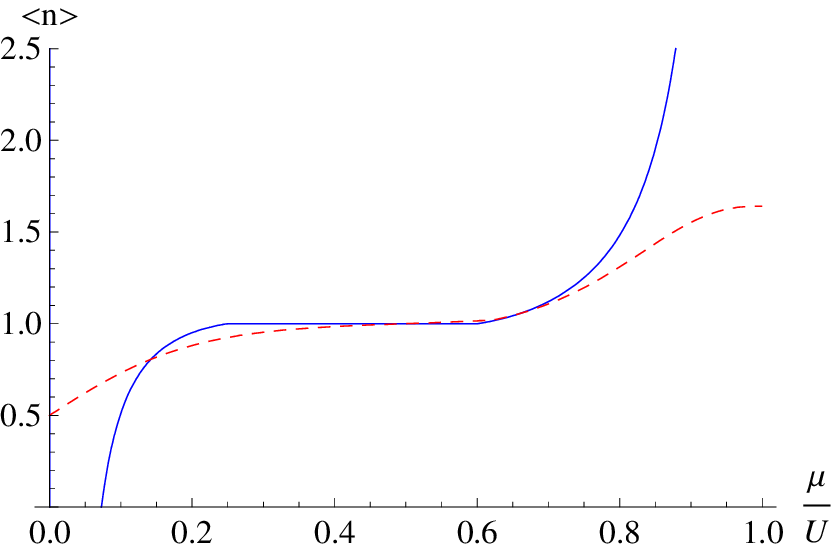}
\caption{Plot of the average number of particles per site as a function of the chemical potential ${\mu}/{U}$ with fixed 
${t}/{U}=0.025$. 
Left shows the mean-field prediction, while right shows the field-theoretic prediction. 
The solid blue curves are ${T}/{U}=0$, and the dashed red curves are ${T}/{U}={0.1}/{k_B}$.} \label{fig3}
\end{center}
\end{figure}
\begin{figure}[h!]
\begin{center}
\includegraphics[width=6cm]{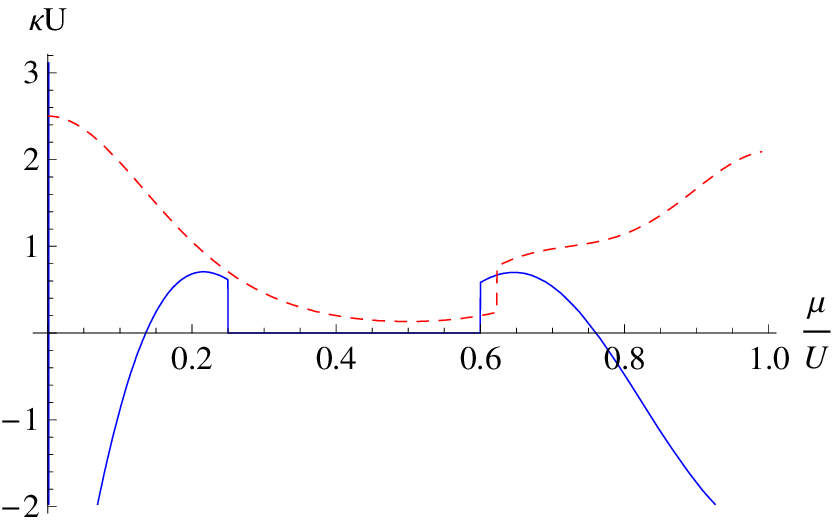}
\includegraphics[width=6cm]{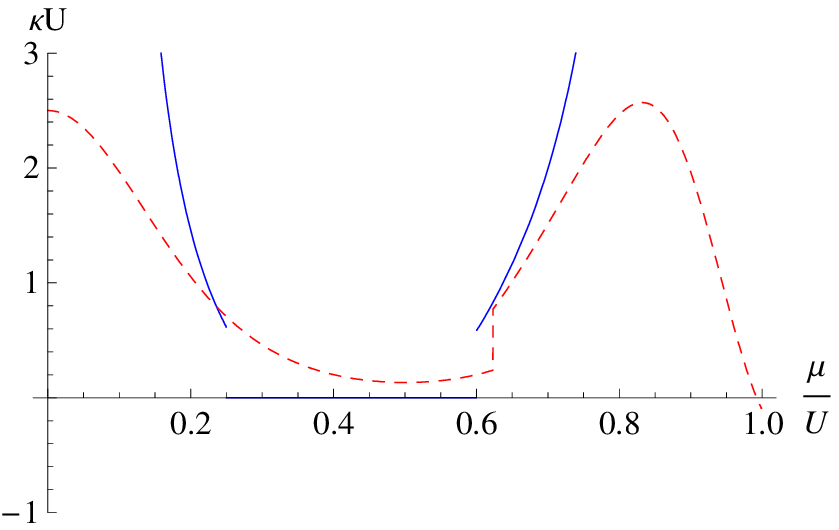}
\caption{Plot of the compressibility ${\kappa}{U}$ as a function of the chemical potential ${\mu}/{U}$ with fixed ${t}/{U}=0.025$. 
Left shows the mean-field prediction, while right shows the field-theoretic prediction. 
The solid blue curves are ${T}/{U}=0$, and the dashed red curves are ${T}/{U}={0.1}/{k_B}$.} \label{fig4}
\end{center}
\end{figure}
Next, we compare the average number of particles per lattice site $\left<n\right>$ as computed in the two theories.
In the ordered phase, the field theoretic prediction for $\left<n\right>$ is given by Eq.~(\ref{eq32}) above, while the mean-field 
result is given by
\begin{equation}
\left<n\right>_{\mathrm{MF}}=-\left.\frac{1}{N_s}
\frac{\partial\mathcal{F}_\mathrm{MF}}{\partial\mu}\right|_{\left|\psi\right|^2=\left|\psi\right|^2_\mathrm{MF}}.
\end{equation}
In the Mott phase, $\left|\psi\right|^2_\mathrm{eq}=\left|\psi\right|^2_\mathrm{MF}=0$, and both theories predict
\begin{equation}
\left<n\right>=\left<n\right>_\mathrm{MF}=-\frac{1}{N}\frac{\partial F_0}{\partial\mu},
\end{equation}
which is simply $\left<n\right>_0$. 
Plots of these two quantities as a function of the chemical potential at a fixed value of the hopping parameter are shown in 
Fig.~\ref{fig3}. 
Although the predictions of both theories agree in the immediate vicinity of the phase boundary, we see that at low temperatures 
$\left<n\right>_\mathrm{MF}$ is not a monotonically increasing function of $\mu$. 
In fact, the plot shows that away from the phase boundary, $\left<n\right>_\mathrm{MF}$ actually decreases with increasing $\mu$, 
i.e. the compressibility is predicted to be negative in the superfluid phase. 
This is directly at odds with the fact that a superfluid has, by definition, a positive compressibility \cite{weichman}. 
The behavior of $\left<n\right>$ as derived from the effective action, on the other hand, fits well with expectation further away 
from the phase boundary. 
This is highlighted in Fig.~\ref{fig4}, which shows the compressibility $\kappa$ for each case.

As a final point of comparison, we specify to the zero temperature case and examine contours in parameter space along which the 
average number density $\left<n\right>$ is constant. 
Inside each Mott lobe, we know that the number density is fixed at the quantum number $n$ of the lobe. In the superfluid phase, for 
fixed $t$ we expect $\left<n\right>$ to increase monotonically with increasing chemical potential. 
This implies that contours of constant $\left<n\right>$ should be monotonic in $\mu$. Such contours for $\left<n\right>=1,2,3$ are 
shown in Fig.~\ref{fig5}. 
Although the mean-field and effective action contours agree close to the lobe tip, the mean-field result exhibits a non-monotonic 
behavior farther away from the lobe tip. 
Given the non-monotonic behavior of $\left<n\right>_\mathrm{MF}$ discussed above, this is not wholly surprising. 
From these considerations, we conclude that the effective action theory has a larger range of validity than the mean field theory 
and, furthermore, we expect an increase in quantitative accuracy when higher powers of $t$ are considered.
\begin{figure}
\begin{center}
\includegraphics[width=6cm]{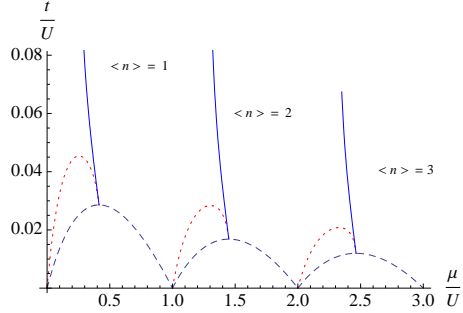}
\caption{Contours of constant $\left<n\right>$ in parameter space for both the mean-field and effective action theories at $T=0$. 
The dotted blue curve shows the first three lobes of the phase boundary. 
The solid blue curves are the predictions of the effective action theory, while the dotted red curves show the predictions of the 
mean-field theory.} \label{fig5}
\end{center}
\end{figure}

\subsection{Effective Action Predictions in the Dynamic Case}
We now turn our attention to the Green's function and the zero-temperature excitation spectra of the Bose-Hubbard model. 
Using Eq.~(\ref{eq22f}), we see that to first order in $t$ the Green's function of the system in the ordered phase can be written as
\begin{equation}
\mathcal{G}(\omega_m,\vec{k})=\frac{a_2^{(0)}(\omega_m)}{1+\frac{2a_4^{(0)}(\omega_{m},0|0,\omega_{m})(a_2^{(0)}(0))^2
\left[6t-\frac{1}{a_2^{(0)}(0)}\right]}{(a_2^{(0)}(\omega_{m}))a_4^{(0)}(0,0|0,0)}-2ta_2^{(0)}(\omega_m)\sum_{\sigma}\cos(k_\sigma d)}, 
\label{eq37}
\end{equation}
where $d$ is the lattice spacing, and $\vec{k}$ is restricted to the first Brillouin zone.
At the phase boundary we have $6t-{1}/{a_2^{(0)}}(0)=0$, and the Green's function reduces to
\begin{equation}
\mathcal{G}(\omega_m,\vec{k})=\frac{a_2^{(0)}(\omega_m)}{1-2ta_2^{(0)}(\omega_m)\sum_\sigma\cos(k_\sigma d)}.
\end{equation}
This is precisely the result obtained in the Mott phase via a resummation of zero loop diagrams \cite{matthias}. 

From Eq.~(\ref{eq18}), we see that the zero-temperature dispersion relation for amplitude excitations $\omega_A(\vec{k})$ satisfies
\begin{equation}
0=\frac{-i}{a_{2\mathrm{R}}^{(0)}(\omega_\mathrm{A})}+
\frac{2a_{4\mathrm{R}}^{(0)}(\omega_\mathrm{A},0|0,
\omega_\mathrm{A})(a_{2\mathrm{R}}^{(0)}(0))^2\left[6t+\frac{i}{a_{2\mathrm{R}}^{(0)}(0)}\right]}{(
a_{2\mathrm{R}}^{(0)}(\omega_\mathrm{A}))^2a_{4\mathrm{R}}^{(0)}(0,0|0,0)}-2t\sum_\sigma\cos\left(k_\sigma d\right) \label{eq40},
\end{equation}
which we recognize also as the condition for poles in Eq.~(\ref{eq37}) continued to real time. 
While too complicated to solve exactly, Eq.~(\ref{eq40}) can be inverted numerically to yield the dispersion relation 
$\omega_\mathrm{A}(\vec{k})$. 
A plot of $\omega_\mathrm{A}(\vec{k})$ taken along the $(1,1,1)$ direction in the first Brillouin zone is shown in Fig.~\ref{fig6}. 
We observe that in the superfluid phase, $\omega_A(\vec{k})$ is gapped and quadratic,
\begin{equation}
\omega_A(\vec{k})\approx \Delta+\eta k^2.
\end{equation} 
Furthermore, at the phase boundary, we find that the dispersion becomes gapless and linear.

Next, we consider the zero-temperature dispersion relation $\omega_\theta(\vec{k})$ of phase excitations. For the Bose-Hubbard model, 
Eq.~(\ref{eq19}) takes the form
\begin{align}
0=&\frac{-i}{a_{2\mathrm{R}}^{(0)}(\omega_\theta)}+\frac{i}{a_{2\mathrm{R}}^{(0)}(0)}-\frac{2a_{2\mathrm{R}}^{(0)}(0)^3
\left(i+6ta_{2R}^{(0)}(0)\right)}{a_{4\mathrm{R}}^{(0)}(0,0|0,0)}\Big[2b(0,0,0,0)
+b(\omega_\theta,-\omega_\theta,0,0)\Big.\nonumber \\
&\Big.+b(0,0,\omega_\theta,-\omega_\theta)-2b(\omega_\theta,0,\omega_\theta,0)-2b(\omega_\theta,0,0,\omega_\theta)\Big]
+2t\left[3-\sum_\sigma\cos\left(k_\sigma d\right)\right].\label{eq41}
\end{align}
We can numerically solve this equation for $\omega_\theta(\vec{k})$. A plot of $\omega_\theta(\vec{k})$ along the $(1,1,1)$ 
direction in the first Brillouin zone is shown in the right of Fig.~\ref{fig6}. 
We see that in the superfluid phase the dispersion is quadratic with 
\begin{equation}
\omega_\theta(\vec{k})\approx\zeta k^2.
\end{equation}
Finally, by comparing Eqs.~(\ref{eq40}) and (\ref{eq41}), we find that at the phase boundary $\omega_A$ and $\omega_\theta$ are 
degenerate.
\begin{figure}
\begin{center}
\includegraphics[width=6cm]{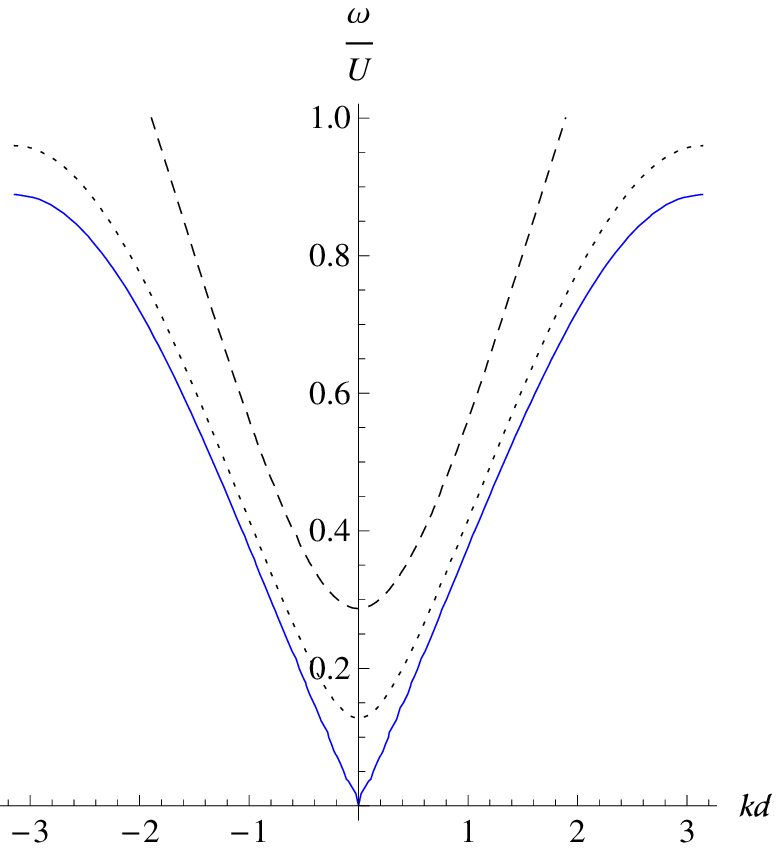}
\hspace{.5cm}
\includegraphics[width=6cm]{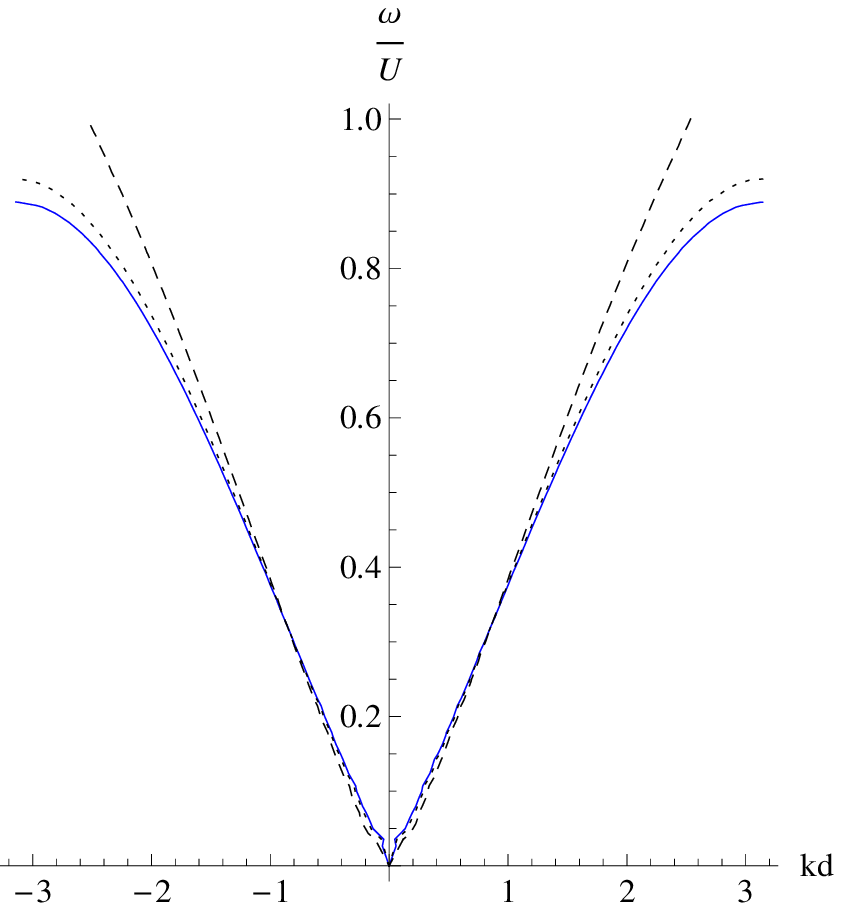}
\caption{Plots of the zero-temperature dispersion relations $\omega_A(\vec{k})$ (Left) and $\omega_\theta(\vec{k})$ (Right) for 
various values of the hopping $t$ with fixed $n=1$, ${\mu}/{U}=\sqrt{2}-1$ and with $\vec{k}=(1,1,1)k/\sqrt{3}$. 
The solid blue lines corresponds to $t=t_c\approx0.028\, U$, which for these values of $\mu$ and $n$ is at the tip of the first 
Mott lobe. 
The dotted yellow lines corresponds to $t=0.03\, U$, and the dashed red lines corresponds to $t=0.035\, U$. Note that amplitude 
excitations exhibit a $t$-dependent energy gap, while the phase excitations are gapless in accordance with Goldstone's theorem.} 
\label{fig6}
\end{center}
\end{figure}
  \begin{figure}
 \begin{center}
\includegraphics[width=6cm]{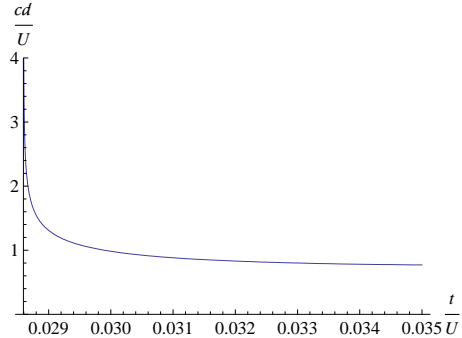}
\caption{Plot of the second sound velocity $c$ as a function of $t/U$ for $t>t_c\approx 0.028\,U$ with 
$n$=1 and ${\mu}/{U}=\sqrt{2}-1$.} \label{fig7}
\end{center}
\end{figure}
Lastly, using the result (\ref{eqss}), we can investigate the behavior of superfluid second sound excitations. 
From our observations above, we find a linear dispersion for small $k$,
\begin{equation}
\omega_s(\vec{k})\approx c k
\end{equation}
with the velocity of sound 
\begin{equation}
c=\sqrt{\Delta\zeta}.
\end{equation}
 Thus, the velocity of second sound at any point near the phase boundary can be found from the above numerical inversions. 
 With this, we plot the sound velocity $c$ as a function of the tunneling $t/U$ in Fig.~\ref{fig7}. We observe that, since near 
the phase boundary a large quadratic term suddenly appears in the phase dispersion relation, the sound velocity jumps immediately 
inside the superfluid phase. This jump must be viewed cautiously, however, as the Ginzburg-Landau expansion is incapable of 
accurately describing critical behavior in the immediate vicinity of the phase 
boundary \cite{kleinert2,zinn-justin,ginzburg,kleinertprl}.  As the mass of the phase excitations begins to increase faster 
than the gap in the amplitude excitations, we see that the sound velocity begins to decrease. Far from the phase boundary, however, 
we know from the seminal Bogoliubov theory that the sound velocity must increase as $\sqrt{t}$ \cite{kleinert,stoof}, confirming 
that our theory is not valid in the deep superfluid phase.

\section{Summary and Conclusion}
In this paper we derived, to first order in the tunneling, the Ginzburg-Landau expansion of the effective action for a very general 
Bosonic lattice Hamiltonian. From the effective action we calculated many static and dynamic system properties of experimental 
interest. In specifying these results to the Bose-Hubbard model, we compared them with the corresponding findings of the 
standard mean-field theory. Although both approaches yield -- up to first order in the tunneling -- the same phase boundary, 
our method gives qualitatively better results deeper in the superfluid phase. Additionally, we were able to find the 
dispersion relation for superfluid excitations, which cannot readily be done in the mean-field approach. The primary 
advantage of our effective-action theory, however, lies in its extensibility. It is straightforward to generalize the 
derivation given in Sections \ref{sec1} and \ref{sec2} by calculating diagrams beyond the tree level in order to include 
higher-order tunneling corrections. As seen in Section \ref{sec4}, this gives a systematic hopping expansion of the self-energy 
function \emph{in both the ordered and non-ordered phases}, providing an arbitrarily precise description of the system dynamics 
near the phase boundary. This should allow, for instance, for the calculation of time-of-flight absorption pictures and their 
corresponding visibilities for the whole phase diagram \cite{hoffmann}. Furthermore, given the generality of the formalism, our 
effective action theory can, in principle, incorporate a variety of interesting effects, such as disordered lattices 
\cite{lewenstein,krutitsky}, vortex dynamics \cite{kleinert}, and tunneling beyond nearest neighbor sites. In particular, an 
effective action for the 
disordered Bose-Hubbard model could give new insight into the nature of the Bose glass phase as a state of short-range order 
\cite{fischer}.

\acknowledgements
The authors thank Hagen Kleinert, Flavio Nogueira, and Matthias Ohliger for fruitful discussions and suggestions. Furthermore, 
we acknowledge financial support from both the German Academic Exchange Service (DAAD) and the German Research Foundation (DFG) 
within the Collaborative Research Center SBF/TR 12 Symmetries and Universality in Mesoscopic Systems.

\begin{thebibliography}{99}
%
\bibitem{boseold7}
A.~J. Legget,
Rev. Mod. Phys. {\bf 73}, 307 (2001).
%
\bibitem{contex}
C.~J. Pethick and H. Smith,
{\it Bose-Einstein Condensation in Dilute Gases}
(Cambridge University Press, Cambridge, 2002).
%
\bibitem{boseold8}
L. Pitaevskii and S. Stringari,
{\it Bose-Einstein Condensation}
(Oxford Science Publications, Oxford, 2003).
%
\bibitem{lewenstein}
M. Lewenstein, A. Sanpera, V. Ahufinger, B. Damski, A. Sen(De), and U. Sen,
Adv. Phys. {\bf 56}, 243 (2007).
%
\bibitem{grimm}
I. Bloch, J. Dalibard, and W. Zwerger,
Rev. Mod. Phys. {\bf 80}, 885 (2008).
%
\bibitem{mandel}
M. Greiner, O. Mandel, T. Esslinger, T. W. H{\"a}nsch, and I. Bloch,
Nature {\bf 415}, 39 (2002).
%
\bibitem{collapse}
M. Greiner, O. Mandel, T.W. H{\"a}nsch, and I. Bloch,
Nature, {\bf 419}, 51 (2002).
%
\bibitem{gerbierpra}
F. Gerbier, A. Widera, S. F{\"o}lling, O. Mandel, T. Gericke, and I Bloch,
Phys. Rev. A  {\bf 72}, 053606 (2005).
%
\bibitem{gerbier}
F. Gerbier, A. Widera, S. F\"olling, O. Mandel, T. Gericke, and I Bloch,
Phys. Rev. Lett. {\bf 95}, 050404 (2005).
%
\bibitem{FW06}
S. F\"olling, A. Widera, T. M\"uller, F. Gerbier, and I. Bloch,
Phys. Rev. Lett. {\bf 97}, 060403 (2006).
%
\bibitem{zurich}
K. G\"unter, T. St\"oferle, H. Moritz, M K\"ohl, and T. Esslinger,
Phys. Rev. Lett. {\bf 96}, 180402 (2006).
%
\bibitem{OOW06}
S. Ospelkaus, C. Ospelkaus, O. Wille, M. Succo, P. Ernst, K. Sengstock, and K. Bongs,
Phys. Rev. Lett. {\bf 96}, 180403 (2006).
%
\bibitem{fischer}
M. P. A. Fisher, P. B. Weichman, G. Grinstein, and D. S. Fisher,
Phys. Rev. B {\bf 40}, 546 (1989). 
%
\bibitem{jaksch}
D. Jaksch,  C. Bruder, J. I. Cirac, C. W. Gardiner, and P. Zoller,
Phys. Rev. Lett. {\bf 81}, 3108 (1998).
%
\bibitem{sachdev}
S. Sachdev,
{\it Quantum Phase Transitions}
(Cambridge University Press, Cambridge, 1999).
%
\bibitem{zoller}
D. Jaksch and P. Zoller,
Ann. Phys. (New York) {\bf 315}, 52 (2005).
%
\bibitem{montecarlo1}
G. G. Batrouni, R. T. Scalettar, and G. T. Zimanyi,
Phys. Rev. Lett. {\bf 65}, 1765 (1990).
%
\bibitem{montecarlo2}
B. Capogrosso-Sansone, S. G. S\"oyler, N. V. Prokof'ev, and B. V. Svistunov,
Phys. Rev. A {\bf 77}, 015602 (2008).
%
\bibitem{montecarlo3}
B. Capogrosso-Sansone, E. Kozik, N.~V. Prokof'ev, and B.~V. Svistunov, 
Phys. Rev. A {\bf 75}, 013619 (2007).
%
\bibitem{monien}
J. K. Freericks and H. Monien,
Phys. Rev. B {\bf 53}, 2691 (1996).
%
\bibitem{monien2}
N. Elstner and H. Monien,
Phys. Rev. B {\bf 59}, 12184 (1999).
%
\bibitem{ednilson}
F.~E.~A. dos Santos and A. Pelster,
Phys. Rev. A  {\bf 79}, 013614 (2009).
%
\bibitem{ziegler1}
K. Ziegler,
Physica A {\bf 208}, 177 (1994). 
%
\bibitem{ziegler2}
K. Ziegler,
J. Low Temp. Phys. {\bf 126}, 1431 (2002).
%
\bibitem{ziegler3}
K. Ziegler,
Las. Phys. {\bf 13}, 587 (2003).
%
\bibitem{krutitsky}
K. V. Krutitsky, A. Pelster, and R. Graham,
New J. Phys. {\bf 8}, 187 (2006).
%
\bibitem{kleinert2}
H. Kleinert and V. Schulte-Frohlinde,
{\it Critical Properties of $\phi^4$-Theories}
(World Scientific, Singapore, 2001).
%
\bibitem{zinn-justin}
J. Zinn-Justin,
{\it Quantum Field Theory and Critical Phenomena}
(Oxford University Press, New York, 2002).
%
\bibitem{peskin}
M. Peskin and D. Schr\"{o}der,
{\it An Introduction to Quantum Field Theory}
(Westview Press, Boulder, 1995).
%
\bibitem{gelfand}
M.~P. Gelfand, R.~R.~P. Singh, and D.~A. Huse,
J. Stat. Phys. {\bf 59}, 1093 (1990).
%
\bibitem{metzner}
W. Metzner,
Phys. Rev. B {\bf 43}, 8549 (1991).
%
\bibitem{matthias}
M. Ohliger,
{\it Dynamics and thermodynamics of spinor bosons in optical lattices},
Diploma Thesis, Free University of Berlin (2008),\\
{\tt  http://users.physik.fu-berlin.de/\~{}ohliger/Diplom.pdf}.
%
\bibitem{fischer2}
M.~E. Fischer, M.~N. Barber, and D. Jasnow,
Phys. Rev. A {\bf 8}, 1111  (1973).
%
\bibitem{roth}
R. Roth and K. Burnett,
Phys. Rev. A {\bf 67}, 031602(R) (2003).
%
\bibitem{hoffmann}
A. Hoffmann and A. Pelster,
eprint: {\tt arXiv:0809.0771}.
%
\bibitem{kleinert}
H. Kleinert,
{\it Multivalued Fields: In Condensed Matter, Electromagnetism, and Gravitation} 
(World Scientific, Singapore, 2008).
%
\bibitem{weichman}
P.~B. Weichman,
Phys. Rev. B {\bf 38}, 8739 (1988).
%
\bibitem{bru}
J.~B.  Bru and T.~C. Dorlas,
J. Stat. Phys. {\bf 113}, 177 (2003).
%
\bibitem{bousante}
P. Buonsante and A. Vezzani,
Phys. Rev. A {\bf 70}, 033608 (2004).
%
\bibitem{krutitsky2}
K. V. Krutitsky, M. Thorwart, R. Egger, and R. Graham,
Phys. Rev. A {\bf 77}, 053609 (2008).  
%
\bibitem{ginzburg}
V.~I. Ginzburg,
Sov. Phys. Solid State {\bf 2}, 1824 (1961).
%
\bibitem{kleinertprl}
H. Kleinert,
Phys. Rev. Lett. {\bf 84}, 286 (2000).
%
\bibitem{stoof}
D. van Oosten,  P. van der Straten, and H. T. C. Stoof,
Phys. Rev. A {\bf 63}, 053601 (2001).
%
\end{thebibliography}
\end{document}